\documentclass[a4paper,11pt]{article}
\pdfoutput=1 

\usepackage{jheppub} 

\usepackage[T1]{fontenc} 

\usepackage{mathrsfs}
\usepackage{amsmath}
\usepackage{braket}
\usepackage{dsfont}
\usepackage{hyperref}

\usepackage{slashed} 
\usepackage{enumitem}
\usepackage{physics}
\usepackage{tikz-feynman}
\usepackage{graphicx}
\usepackage{dcolumn}
\usepackage{bm}

\usepackage[force]{feynmp-auto} 
\usepackage{floatrow}
\newfloatcommand{capbtabbox}{table}[][\FBwidth]
\usepackage{subfig}

\def\p{\textbf{p}}

\def\0{\textbf{0}}
\def\P{\textbf{P}}
\def\J{\textbf{J}}
\def\K{\textbf{K}}
\def\W{\textbf{W}}
\def\x{\textbf{x}}

\renewcommand\thefigure{\arabic{figure}}

\allowdisplaybreaks

\title{\boldmath Pseudo-Hermitian QFT: relativistic scattering and symmetry structure}
\author[a]{Ruifeng Leng,}
\affiliation[a]{Department of Physics and Center for Field Theory and Particle Physics, \\Fudan University, Shanghai 200438, China}
\emailAdd{lruifeng@fudan.edu.cn} 
\author[b]{Cheng-Yang Lee,}
\affiliation[b]{Department of Physics and Chongqing Key Laboratory for Strongly Coupled Physics,
Chongqing University, Chongqing 401331, China} 
\emailAdd{chengyanglee@outlook.com} 
\author[b]{Siyi Zhou}
\emailAdd{siyi@cqu.edu.cn}

\abstract{Unitarity is a cornerstone of quantum theory, ensuring the conservation of probability and information. Although non-Hermitian Hamiltonians are typically associated with open or dissipative systems, pseudo-Hermitian quantum mechanics shows that real spectra and unitary evolution can still emerge through a suitably defined inner product. Motivated by this insight, we extend the pseudo-Hermitian framework to relativistic quantum field theory and construct a consistent formulation of scattering processes. 
A novel structural feature of this theory is the presence of distinct metric operators for the in and out sectors, connected through a nontrivial metric projector that guarantees global probability conservation under pseudo-unitary time evolution. 
We further develop a general symmetry formalism, showing that each symmetry generally corresponds to two pseudo-unitary operators associated with the in and out metrics, respectively. 
Within this framework, the scattering matrix admits a perturbative expansion through the Dyson series and remains Lorentz invariant and unitary, remarkably in complete agreement with the conventional Hermitian case. 
The fundamental $CPT$ theorem is also shown to hold. 
Our results provide a rigorous foundation for interacting pseudo-Hermitian quantum field theories and open new directions for exploring their possible physical implications beyond the standard Hermitian paradigm.}

\begin{document} 
\maketitle
\flushbottom

\section{Introduction} 

In particle physics, unitarity has played an indispensable role in the development of the Standard Model (SM), guiding the theoretical framework that led to the discovery of the Higgs boson~\cite{Higgs:1964pj,Higgs:1964ia,Englert:1964et,Guralnik:1964eu}. It has also served as a guiding principle in addressing the black hole information paradox and has inspired the development of holography~\cite{tHooft:1993dmi,Susskind:1994vu} and the AdS/CFT correspondence~\cite{Maldacena:1997re}.



Unitarity ensures the conservation of probability and information in quantum theory. 
Conversely, non-unitary theories, often characterized by non-Hermitian Hamiltonians, typically describe open or dissipative systems~\cite{Moiseyev_2011,Ashida:2020dkc}. 
However, it is possible to construct a new positive-definite inner product that depends on the dynamics of the system and restores unitarity and resolving the related conceptual challenges~\cite{Das:2009it,Das:2011zza}.
Although a general pseudo-Hermitian Hamiltonian may possess complex eigenvalues, the converse is not true: a Hamiltonian with a real spectrum does not necessarily need to be Hermitian. This important insight originates from the pioneering studies by Bender and Boettcher on $PT$ symmetric Hamiltonians~\cite{Bender:1998ke,Bender:2007nj}, and by Mostafazadeh on pseudo-Hermitian systems~\cite{Mostafazadeh:2001jk,Mostafazadeh:2001nr,Mostafazadeh:2002id,Mostafazadeh:2008pw}. These works demonstrate that pseudo-Hermitian Hamiltonians can exhibit entirely real spectra under appropriate conditions.
This naturally leads to two fundamental questions: \textit{Can such pseudo-Hermitian Hamiltonians describe closed systems? What happens to unitarity?} 
In this work, we address these questions within the framework of pseudo-Hermitian quantum field theory (QFT). Our study contributes to the ongoing effort to broaden the scope of quantum theory beyond the traditional requirement of Hermiticity.

Pseudo-Hermitian QFT has been explored across a variety of domains, including condensed matter physics~\cite{Kapit:2008dp,LeClair_2007}, particle physics~\cite{Mason:2023sgr,Ahluwalia:2022zrm,Lee:2023aip,Sablevice:2023odu,Ahluwalia:2023slc}, and dS/CFT correspondence~\cite{Fei:2015kta,Sato:2015tta,Anninos:2011ui,Ng:2012xp,Chang:2013afa,Ryu:2022ffg}. 
Most studies so far have focused on the free theories that remain local, preserve Lorentz invariance, and are governed by positive definite Hermitian Hamiltonians.
To address the above questions, we must extend the framework to interacting theories. In this paper, we formulate the relativistic scattering theory for pseudo-Hermitian QFTs. A necessary condition for studying scattering processes is that the spectrum of the pseudo-Hermitian Hamiltonian be real. Under this condition, the time evolution of the in and out states is well defined and coincides with that of free states in the asymptotic past and future. The requirement that the in and out states continuously evolve into the free states determines a specific inner product in each sector, both orthogonal and positive definite. To connect these two sectors, we introduce a nontrivial structural element --- the metric projector, which consistently mediates their relation. The resulting $S$-matrix, defined within this framework, admits a perturbative expansion given by the Dyson series and remains Lorentz invariant as well as unitary, perhaps surprisingly in complete agreement with the Hermitian case. 
A general symmetry formalism naturally follows from this construction, revealing the distinctive structural features of the pseudo-Hermitian framework that set it apart from conventional Hermitian theory.

This paper is organized as follows. In section~\ref{sec:pH_QM}, we begin by reviewing the fundamental building blocks of pseudo-Hermitian quantum mechanics and argue that the essential physical behavior arises from the pseudo-Hermitian structure itself.
In section~\ref{sec:scatter}, we establish the formulation of relativistic scattering in the pseudo-Hermitian QFT and analyze how pseudo-Hermiticity modifies the conventional construction of the $S$-matrix, the central quantity from which observables such as cross sections and decay rates are obtained. Finally, in section~\ref{sec:scatter_symmetry}, we develop a general symmetry formalism and discuss how pseudo-Hermiticity challenges the conventional formalism and associated symmetry conditions.


\section{Pseudo-Hermitian quantum mechanics}
\label{sec:pH_QM}

In this section, we introduce the concept of pseudo-Hermitian operators and derive the basic properties of pseudo-Hermitian Hamiltonians~\cite{Mostafazadeh:2008pw}. We adapt this framework to the formalism commonly used in QFT, which provides the theoretical foundation for the scattering theory developed in the subsequent sections.

Let $\mathscr{H}$ be a Hilbert space equipped with a Hermitian linear automorphism $\eta$
\begin{equation}
\eta^{\dag} = \eta \, .
\label{eq:eta_defn_0}
\end{equation}
Given this structure, the $\eta$-pseudo-Hermitian conjugation of a linear operator $O:\mathscr{H} \rightarrow \mathscr{H}$ is defined by
\begin{equation}
O^{\#}\equiv\eta^{-1}O^{\dag}\eta \, .
\label{eq:pseudo_H_conj}
\end{equation}
From this definition and the Hermiticity of $\eta$~\eqref{eq:eta_defn_0}, it follows that
\begin{equation}
O^{\# \#} = O \, .
\end{equation}
In particular, when $\eta = \boldsymbol{1}$,~\footnote{To be clear, `$\boldsymbol{1}$' denotes the identity operator on the Hilbert space.} the pseudo-Hermitian conjugation reduces to the standard Hermitian conjugation.
An operator $O$ is said to be pseudo-Hermitian if it is invariant under the pseudo-Hermitian conjugation:
\begin{equation}
O^{\#} = O \, .
\label{eq:pseudo_H_0}
\end{equation}
While we are quite familiar with quantum mechanical systems governed by Hermitian Hamiltonians, our focus in this work is on systems whose Hamiltonians are pseudo-Hermitian, i.e.,
\begin{equation}
H^{\#} = \eta^{-1}H^{\dag}\eta = H \, .
\label{eq:pseudo_H_1}
\end{equation}
It is important to emphasize that the time evolution operator~\footnote{For simplicity, here we restrict our attention to a time-independent Hamiltonian $H$. 
In the more general case, where a quantum system is governed by a possibly pseudo-Hermitian and time-dependent
Hamiltonian $H(t)$, the non-conservation of the Dirac-inner product over time can be derived from the Schr\"{o}dinger equation together with the pseudo-Hermitian condition~\eqref{eq:pseudo_H_1}.}
\begin{equation}
U(t) = e^{-iHt} \, ,
\end{equation}
is generally not unitary in the conventional sense, since $H$ is generally not Hermitian. As a result, the Dirac-inner product between two arbitrary states evolves as
\begin{equation}
\langle \phi (t) |\psi (t) \rangle 
=\langle \phi (0)| e^{iH^{\dag}t} e^{-iHt} |\psi (0) \rangle 
\neq \langle \phi (0) |\psi (0) \rangle 
\, ,
\end{equation}
and is therefore not conserved over time. This apparent non-unitarity would imply a violation of probability conservation under time evolution.
An immediate solution to this issue is to define a modified inner product with respect to which the time evolution becomes unitary. This leads to the so-called $\eta$-inner product~\cite{Mostafazadeh:2001jk}:
\begin{equation}
\langle \phi|\psi \rangle_{\eta} \equiv \langle \phi | \eta|\psi \rangle\ 
\, ,
\quad
\forall |\phi \rangle, |\psi \rangle \in \mathscr{H} \, ,
\label{eq:eta_prod_1}
\end{equation}
where $\eta$ is sometimes called a metric operator.
This quadratic form reduces to the standard Dirac-inner product when $\eta=\boldsymbol{1}$.
Within this definition, time evolution would be unitary with respect to the $\eta$-inner product:~\footnote{For convenience, we denote $|\phi (0) \rangle \equiv |\phi \rangle$ and $|\psi (0) \rangle \equiv |\psi \rangle$.}
\begin{align}
\langle \phi (t) |\psi (t) \rangle_{\eta} 
=\langle \phi (0)| e^{iH^{\dag}t} \eta e^{-iHt} |\psi (0) \rangle 
=\langle \phi (0)| \eta e^{iHt} e^{-iHt} |\psi (0) \rangle 
= \langle \phi (0) |\psi (0) \rangle_{\eta} 
\, ,
\label{eq:evo_eta_inner_1}
\end{align}
where we have imposed the pseudo-Hermitian condition of $H$~\eqref{eq:pseudo_H_1}.
Besides, analogous to the Hermitian conjugation defined as the adjoint with respect to the Dirac-inner product, the pseudo-Hermitian conjugation can be interpreted as the adjoint with respect to the $\eta$-inner product.

A natural question arises whether a pseudo-Hermitian system can be defined in such a way that it supports both a probabilistic interpretation and unitary time evolution. The essential problem is that the $\eta$-inner product is not necessarily positive-definite in the original definition. Thus, a general $\eta$-inner product may fail to provide a physically meaningful probabilistic interpretation.
It is important to note, however, that a probabilistic description does not strictly require $\eta$ to be positive-definite. When $\eta$ is positive-definite, commonly denoted by $\hat{\eta}$, the corresponding $\hat{\eta}$-inner product is strictly positive for all nonzero vectors and vanishes only for the zero vector.
Fortunately, it is generally possible to construct a canonical Hermitian operator $\eta_{0}$ from a given metric operator $\eta$~\cite{Das:2011zza}, such that 
\begin{equation}
H^{\#} = \eta^{-1}H^{\dag}\eta = \eta_{0}^{-1}H^{\dag}\eta_{0} = H \, ,
\label{eq:pseudo_H_2}
\end{equation}
which implies the Hamiltonian $H$ is also pseudo-Hermitian with respect to this $\eta_{0}$-inner product. Moreover, time evolution maintains unitary under the $\eta_{0}$-inner product:
\begin{equation}
\langle \phi|\psi \rangle_{\eta_{0}} \equiv \langle \phi | \eta_{0} |\psi \rangle\ 
\, ,
\quad
\forall |\phi \rangle, |\psi \rangle \in \mathscr{H} \, .
\label{eq:eta_prod_0}
\end{equation}
Therefore, this carefully constructed operator $\eta_{0}$ ensures a consistent quantum mechanical description. 
An important distinction between pseudo-Hermitian and Hermitian systems lies in the nature of the inner product. In Hermitian systems, the inner product is fixed and does not depend on the dynamics of the system. In contrast, in a pseudo-Hermitian system, the $\eta_{0}$-inner product nontrivially depends on the Hamiltonian.
In particular, $\eta_{0}$-inner product is non-negative and vanishes not only for the zero vector but also for any eigenstate of the pseudo-Hermitian Hamiltonian $H$ associated with a complex eigenvalue. 
Since the spectrum of a pseudo-Hermitian Hamiltonian can be either entirely real or consist of complex conjugate pairs~\cite{Mostafazadeh:2001jk}, the specific metric operator $\eta_{0}$ is not generally positive-definite.
In fact, a positive-definite metric operator $\hat{\eta}$ exists if and only if the pseudo-Hermitian Hamiltonian $H$ has a real spectrum~\cite{Mostafazadeh:2002id}. In this case, the $\eta_{0}$ constructed in ref.~\cite{Das:2011zza} reduces to $\hat{\eta}$. 
It is worth noting that Hermiticity is a sufficient but not necessary condition for a real spectrum, while pseudo-Hermiticity is a necessary but not sufficient condition for spectral reality~\cite{Mostafazadeh:2001nr}.

\subsection{Pseudo-unitary time evolution}
\label{sec:pic_T}

All predictions in quantum theory are generally formulated in terms of inner products between bra and ket vectors or as matrix elements of observable operators. The time evolution of these quantities can be described through several equivalent frameworks, commonly referred to as quantum pictures. Since we have introduced a canonical metric operator $\eta_{0}$~\eqref{eq:eta_prod_0} to redefine the physical inner product in a pseudo-Hermitian quantum theory, it is essential to revisit time evolution in the two commonly used pictures and reformulate each consistently with the canonical metric operator $\eta_{0}$.

Without loss of generality, we take $t_{0}$ as the reference time at which all pictures are aligned, i.e.,
\begin{align}
O(t_{0}) &\equiv O_{S}(t_{0}) = O_{H}(t_{0}) = O_{I}(t_{0}) \, , 
\label{eq:align_pic_01} \\
\ket{\psi(t_{0})} &\equiv \ket{\psi_{S}(t_{0})} = \ket{\psi_{H}(t_{0})} = \ket{\psi_{I}(t_{0})} \, , 
\quad
\forall |\psi (t_{0})\rangle \in \mathscr{H} \, .
\label{eq:align_pic_02}
\end{align}
To avoid confusion, in this section, we shall systematically assign the subscripts $S$, $H$, and $I$ to the states and operators to denote Schr\"{o}dinger, Heisenberg, and interaction pictures, respectively. Although this alignment is also valid for other pictures, we focus on these two standard cases here.
The matrix element of an observable operator is defined with respect to the canonical $\eta_{0}$-inner product as
\begin{equation}
\langle \phi(t)| O(t) |\psi(t) \rangle_{\eta_{0}} \equiv \langle \phi(t) | \eta_{0} O(t) |\psi(t) \rangle
\, , 
\label{eq:obs_mat_eta_1}
\end{equation}
and remains invariant under different choices of time evolution pictures. As a result, the picture label can be omitted in this context.~\footnote{Equivalently, one may evaluate and label it in any specific picture, as we will do later.} This invariance reflects the fundamental principle that the evolution of physical quantities should be independent of the chosen picture. In other words, this matrix element can be computed equivalently in all pictures, and this invariance determines the relationships between the specific forms of state and operator evolution in each picture, providing the corresponding transformation rules among them.
In addition, requiring that the expectation value of an observable~\eqref{eq:obs_mat_eta_1} be real for any state leads to the condition
\begin{align}
\langle \psi(t)| O(t) |\psi(t) \rangle_{\eta_{0}}^{*} 
= \langle \psi(t)| O^{\dag}(t) \eta_{0} |\psi(t) \rangle
= \langle \psi(t)| O(t) |\psi(t) \rangle_{\eta_{0}}
\, ,
\end{align}
which implies that the observable operator must be pseudo-Hermitian in all pictures,~\footnote{An observable operator is allowed to be both pseudo-Hermitian and Hermitian if it satisfies both definitions simultaneously.} i.e., 
\begin{equation}
O^{\#}(t) = \eta_{0}^{-1}O^{\dag}(t)\eta_{0} = O(t) \, .
\label{eq:pseudo_O_1}
\end{equation}

\noindent\textbf{Schr\"{o}dinger picture} \ \  
The operators representing observables are typically taken to be time-independent, while the entire time dependence of the system is encoded in the quantum state. The evolution of a state is governed by the Schr\"{o}dinger equation, which reveals the reason why this framework is called the Schr\"{o}dinger picture. 
Nevertheless, in the most general setting, an observable operator may possess explicit time dependence. To account for this possibility, we adopt the following formulation for time evolution:
\begin{align}
O(t) = O_{S}(t) \, , 
\quad
\ket{\psi(t)} = \ket{\psi_{S}(t)} \equiv U_{S}(t, t_{0}) \ket{\psi_{S}(t_{0})} \, ,
\label{eq:evo_S_1}
\end{align}
where $O_{S}(t)$ denotes the observable operator, which possibly depend on time explicitly in the Schr\"{o}dinger picture.
$\ket{\psi_{S}(t)}$  is the quantum state at the instant $t$, obtained by evolving the initial state $\ket{\psi_{S}(t_{0})}$ from the initial time $t_{0}$, using the time evolution operator
$U_{S}(t, t_{0})$ in the Schr\"{o}dinger picture. This evolution is typically generated by the full Hamiltonian $H$ of the system via the Schr\"{o}dinger equation:
\begin{align}
i \partial_{t} \ket{\psi_{S}(t)} = H \ket{\psi_{S}(t)} \, ,
\label{eq:evo_state_S_1}
\end{align}
which leads to the differential equation for the time evolution operator $U_{S}(t, t_{0})$
\begin{align}
i \partial_{t} U_{S}(t, t_{0}) = H \, U_{S}(t, t_{0}) \, ,
\label{eq:evo_U_S_1}
\end{align}
whose general solution is given by an exponential
\begin{align}
U_{S}(t, t_{0}) = e^{-i H (t - t_{0})} \, .
\label{eq:evo_U_S_2}
\end{align}
It is important to note that the Hamiltonian $H$ is not required to be Hermitian necessarily here, but pseudo-Hermitian~\eqref{eq:pseudo_H_2} instead. 
As a result, the time evolution operator $U_{S}(t, t_{0})$ is not necessarily unitary with respect to the Dirac-inner product, i.e.,
\begin{align}
U_{S}^{\dag}(t, t_{0}) = e^{i H^{\dag} (t - t_{0})} \neq U_{S}^{-1}(t, t_{0}) \, ,
\end{align}
even though we retain the standard notation for consistency with standard quantum theory.
However, by taking into account the pseudo-Hermiticity of $H$~\eqref{eq:pseudo_H_2}, the evolution operator $U_{S}(t, t_{0})$ satisfies the pseudo-unitarity condition with respect to the metric operator $\eta_{0}$~\footnote{The conditions for pseudo-unitarity or pseudo-antiunitarity both take the form $\eta_{0}^{-1} U^{\dag} \eta_{0} = U^{-1}$.} 
\begin{align}
U_{S}^{\dag}(t, t_{0}) \eta_{0} = \eta_{0} e^{i H (t - t_{0})} = \eta_{0} U_{S}^{-1}(t, t_{0}) \, ,
\label{eq:pseudo_U_S_1}
\end{align}
so that the evolution of the $\eta_{0}$-inner product remains unitary as we have shown in eq.~\eqref{eq:evo_eta_inner_1}.
The $\eta_{0}$-matrix element of an observable operator can thus be rewritten as
\begin{align}
\langle \phi(t)| O(t) |\psi(t) \rangle_{\eta_{0}} 
= \langle \phi(t_{0}) | U_{S}^{-1}(t, t_{0}) O_{S}(t) U_{S}(t, t_{0}) |\psi(t_{0}) \rangle_{\eta_{0}}
\, ,
\label{eq:obs_mat_eta_S_1}
\end{align}
whose time evolution equation can be derived by applying that of the operator $O(t)$~\eqref{eq:evo_S_1} and the quantum state $\ket{\psi_{S}(t)}$~\eqref{eq:evo_state_S_1}:
\begin{align}
\dfrac{d}{dt} \langle \phi(t)| O(t) |\psi(t) \rangle_{\eta_{0}} 
= \langle \phi_{S}(t) | -i \left[O_{S}(t), H \right] + \dot{O}_{S}(t) |\psi_{S}(t) \rangle_{\eta_{0}} 
\, ,
\label{eq:evo_mat_S_1}
\end{align}
where we have inserted the pseudo-unitarity of $U_{S}(t, t_{0})$~\eqref{eq:pseudo_U_S_1} in the second step. 
Nevertheless, an especially important case arises when the system is conservative, meaning the matrix element of an observable remains constant in time. In this case, the vanishing of the time derivative in eq.~\eqref{eq:evo_mat_S_1} leads to the conservation condition:
\begin{align}
-i \left[O_{S}(t), H \right] + \dot{O}_{S}(t) = 0
\, ,
\label{eq:con_O_S_1}
\end{align}
which further reduces to the commutation relation
\begin{align}
\left[O_{S}, H \right] = 0 \, ,
\label{eq:con_O_S_2}
\end{align}
if $O_{S}(t) = O_{S} = O(t_{0})$ is time independent.
This formulation naturally motivates the transition to the Heisenberg picture, in which the time dependence is transferred entirely to the operators, while the quantum states remain fixed. This shift in perspective offers a complementary view of quantum dynamics, often more convenient for identifying conserved quantities and symmetry generators.

\noindent\textbf{Heisenberg picture} \ \  
The quantum state is held fixed, while the time dependence is entirely transferred to the observable operators. The time evolution is expressed as:
\begin{align}
O(t) = O_{H}(t) \equiv U_{S}^{-1}(t, t_{0}) O_{S}(t) U_{S}(t, t_{0}) \, , 
\quad
\ket{\psi(t)} = \ket{\psi_{H}} \equiv \ket{\psi_{H}(t_{0})} = \ket{\psi(t_{0})} \, ,
\label{eq:evo_H_1}
\end{align}
where $O_{H}(t)$ denotes the observable operator at the instant $t$ in the Heisenberg picture, obtained by evolving the Schr\"{o}dinger-picture operator $O_{S}(t)$ via the pseudo-unitarity time evolution operator $U_{S}(t, t_{0})$~\eqref{eq:evo_U_S_2}. 
The quantum state $\ket{\psi_{H}}$ remains constant in time and coincides with $\ket{\psi_{S}(t_{0})}$~\eqref{eq:align_pic_02} at the reference time $t_{0}$.

Under the Heisenberg evolution framework in eq.~\eqref{eq:evo_H_1}, one can immediately verify that the $\eta_{0}$-matrix element $\langle \phi_{H}| O_{H}(t) |\psi_{H} \rangle_{\eta_{0}}$ reproduces the corresponding expression in the Schr\"{o}dinger picture in eq.~\eqref{eq:obs_mat_eta_S_1}.
Thus, the time evolution equation for $O_{H}(t)$~\eqref{eq:evo_H_1}, the well-known Heisenberg equation, follows directly from eq.~\eqref{eq:evo_mat_S_1}
\begin{align}
\dot{O}_{H}(t) &= U_{S}^{-1}(t, t_{0}) \left( -i \left[O_{S}(t), H \right] + \dot{O}_{S}(t) \right) U_{S}(t, t_{0})
\nonumber \\
&= -i \left[O_{H}(t), H \right] + \partial_{t} O_{H}(t)
\, ,
\label{eq:evo_U_H_1}
\end{align}
with 
\begin{align}
\partial_{t} O_{H}(t) \equiv U_{S}^{-1}(t, t_{0}) \dot{O}_{S}(t) U_{S}(t, t_{0})
\, ,
\end{align}
denoting the Heisenberg-picture counterpart of the explicit time dependence $\dot{O}_{S}(t)$.~\footnote{Of course, eq.~\eqref{eq:evo_U_H_1} can also be obtained by directly calculating $\dot{O}_{H}(t)$ using the Schr\"{o}dinger equation for the evolution operator $U_{S}(t, t_{0})$~\eqref{eq:evo_U_S_1}.}
The conservation condition in eq.~\eqref{eq:con_O_S_1} then translates into the Heisenberg picture as
\begin{align}
\dot{O}_{H}(t) = 0 \, ,
\label{eq:con_O_H_1}
\end{align}
which implies that $O_{H}(t) = O_{H} = O(t_{0})$ is a constant operator in time.
In particular, if $O_{H}(t)$ does not depend on time explicitly [i.e., $\partial_{t} O_{H}(t) = 0$], the Heisenberg equation for $O_{H}(t)$~\eqref{eq:evo_U_H_1} simplifies to
\begin{align}
\dot{O}_{H}(t) = -i \left[O_{H}(t), H \right] 
\, ,
\label{eq:evo_U_H_2}
\end{align}
so that the conservation condition for $O_{H}(t)$ given in eq.~\eqref{eq:con_O_H_1} implies the commutation relation
\begin{align}
\left[O_{H}, H \right] = 0 \, ,
\label{eq:con_O_H_2}
\end{align}
consistent with the requirement~\eqref{eq:con_O_S_2} derived in the Schr\"{o}dinger picture, due to the alignment of the two pictures at the reference time $t_{0}$ as specified in eq.~\eqref{eq:align_pic_01}.
In the following sections of this paper, we will stay in the Heisenberg picture unless stated otherwise, as it provides a more convenient framework for discussing time-dependent quantities and causality. For notational simplicity, the explicit subscript $H$ will be omitted.


\subsection{Generalized representation of the Poincar\'{e} group}
\label{sec:rep_Poin_pH}

Much of the structure of a symmetry described by a Lie group is encoded in the behavior of its elements near the identity, which is captured by its Lie algebra.
For the Poincar\'{e} group, its Lie algebra consists of $6+4=10$ generators, following the commutation rules:
\begin{align}
[J^{\mu\nu}, J^{\rho\sigma}] 
&= i \left(g^{\nu\rho} J^{\mu\sigma} - g^{\mu\rho} J^{\nu\sigma} 
- g^{\sigma\mu} J^{\rho\nu} + g^{\sigma\nu} J^{\rho\mu} \right) \, , 
\label{eq:poin_lie_11} \\
[P^{\mu}, J^{\rho\sigma}] 
&= i \left(g^{\mu\rho} P^{\sigma} - g^{\mu\sigma} P^{\rho} \right) \, , 
\label{eq:poin_lie_12} \\
[P^{\mu}, P^{\rho}] &= 0 \, ,
\label{eq:poin_lie_13}
\end{align}
where $J^{\rho\sigma}$ is antisymmetric, i.e., $J^{\rho\sigma}=-J^{\sigma\rho}$.
In this algebra, we identify the operators $P^{1}$, $P^{2}$, $P^{3}$ as the components of the momentum vector; $J^{23}$, $J^{31}$, $J^{12}$ form the components of the angular momentum vector; and $P^{0}$ represents the energy operator, i.e., the Hamiltonian $H$.

In quantum theory, conserved operators always play a crucial role. As discussed in section~\ref{sec:pic_T}, 
any operator that is conserved and does not depend explicitly on time must commute with the Hamiltonian $P^{0}=H$.
Based on eqs.~\eqref{eq:poin_lie_11}-\eqref{eq:poin_lie_13}, these operators are the three-momentum vector
\begin{align}
\P = \left\{ P^{1}, P^{2}, P^{3} \right\} \, , 
\end{align}
and the angular momentum vector~\footnote{Alternatively, the angular momentum vector can be defined via the relations $J^{ij} \equiv -i \left[ J^{i}, J^{j} \right]  = \epsilon^{ijk} J^{k}$, so that $J^{k} = \dfrac{1}{2} \epsilon_{kij} J^{ij}$.}
\begin{align}
\J = \left\{ J^{23}, J^{31}, J^{12} \right\} \, , 
\end{align}
and of course the Hamiltonian $P^{0} = H$ itself.
The remaining generators, which form the Lorentz boost vector,
\begin{align}
\K = \left\{ J^{01}, J^{02}, J^{03} \right\} \, , 
\end{align}
are also conserved but depend on time explicitly.
In this three dimensional notation, we rewrite the Lie algebra of the Poincar\'{e} group given in eqs.~\eqref{eq:poin_lie_11}-\eqref{eq:poin_lie_13} as
\begin{align}
[ J^{i}, J^{j} ] &= i \epsilon_{ijk} J^{k} \, , 
\label{eq:poin_lie_21} \\
[ J^{i}, K^{j} ] &= i \epsilon_{ijk} K^{k} \, , 
\label{eq:poin_lie_22} \\
[ K^{i}, K^{j} ] &= -i \epsilon_{ijk} J^{k} \, , 
\label{eq:poin_lie_23} \\
[ J^{i}, P^{j} ] &= i \epsilon_{ijk} P^{k} \, , 
\label{eq:poin_lie_24} \\
[ K^{i}, P^{j} ] &= -i H \delta_{ij} \, , 
\label{eq:poin_lie_25} \\
[ J^{i}, H ] &= [ P^{i}, H ] = [ H, H ] = 0 \, , 
\label{eq:poin_lie_26} \\
[ K^{i}, H ] &= -i P^{i} \, ,
\label{eq:poin_lie_27} 
\end{align}
which corresponds to the Hermitian conjugate form
\begin{align}
[ J^{i}, J^{j} ] &= i \epsilon_{ijk} J^{k} \, , 
\label{eq:poin_lie_21_h} \\
[ J^{i}, K^{j \dag} ] &= i \epsilon_{ijk} K^{k \dag} \, , 
\label{eq:poin_lie_22_h} \\
[ K^{i \dag}, K^{j \dag} ] &= -i \epsilon_{ijk} J^{k} \, , 
\label{eq:poin_lie_23_h} \\
[ J^{i}, P^{j} ] &= i \epsilon_{ijk} P^{k} \, , 
\label{eq:poin_lie_24_h} \\
[ K^{i \dag}, P^{j} ] &= -i H^{\dag} \delta_{ij} \, , 
\label{eq:poin_lie_25_h} \\
[ J^{i}, H^{\dag} ] &= [ P^{i}, H^{\dag} ] = [ H^{\dag}, H^{\dag} ] = 0 \, , 
\label{eq:poin_lie_26_h} \\
[ K^{i \dag}, H^{\dag} ] &= -i P^{i} \, .
\label{eq:poin_lie_27_h} 
\end{align}
As in the standard Hermitian system, we have already required the momentum $\P$ and angular momentum $\J$ to be Hermitian:
\begin{align}
\P = \P^{\dag} \, , 
\quad
\J = \J^{\dag} \, .
\end{align}
However, in order to adapt to a pseudo-Hermitian Hamiltonian $H$~\eqref{eq:pseudo_H_2},
it is intriguing that the Lorentz boost $\K$ is not necessarily Hermitian, but rather must be $\eta_{0}$-pseudo-Hermitian
\begin{align}
\K^{\#} = \eta_{0}^{-1}\K^{\dag}\eta_{0} = \K \, .
\label{eq:pseudo_K_1} 
\end{align}
Meanwhile, the momentum $\P$ and angular momentum $\J$ are constrained to be both pseudo-Hermitian and Hermitian simultaneously
\begin{alignat}{2}
\P &= \P^{\dag} \, , 
\quad
&\P^{\#}& = \eta_{0}^{-1}\P^{\dag}\eta_{0} = \P \, ,
\label{eq:pseudo_P_1} \\
\J &= \J^{\dag} \, ,
\quad
&\J^{\#}& = \eta_{0}^{-1}\J^{\dag}\eta_{0} = \J \, ,
\label{eq:pseudo_J_1} 
\end{alignat}
so that $\P$ and $\J$ share a common complete set of physical eigenstates and possess real spectra, thereby enabling the decomposition of the Hilbert space as usual.~\footnote{More precisely, quantum states are labeled by the eigenvalues of the two Casimir invariants of the Poincar\'{e} group: $P_{\mu} P^{\mu}$ and $W_{\mu} W^{\mu}$, which correspond to the mass and spin of the particle respectively.
$P^{\mu}$ is the four-momentum operator, and $W^{\mu} \equiv \frac{1}{2} \epsilon^{\mu \nu \rho \sigma} P_{\nu} J_{\rho \sigma}$ is the Pauli-Lubanski vector.}

\subsection{Complete biorthonormal eigenbasis \& completeness relation}
\label{sec:bi_basis_H}

The results that we have presented in the previous sections so far apply generally to all pseudo-Hermitian Hamiltonians.
However, to formulate a well-defined scattering theory, it is essential that we require the `in' and `out' states to smoothly evolve into effective free-particle states, i.e., asymptotic states, in the asymptotic time limits $t\rightarrow\pm\infty$. This asymptotic behavior is incompatible with a complex energy spectrum, as will be discussed in detail in section~\ref{sec:in_out}. For this reason, we restrict our attention to pseudo-Hermitian Hamiltonians with entirely real spectra throughout the rest of this paper.
Moreover, the presence of the \textit{exact} $PT$ symmetry guarantees the reality of the spectrum~\cite{Dorey:2001uw,Dorey:2003sq,Mostafazadeh:2002id}, and therefore offers a powerful framework for constructing physically admissible scattering Hamiltonians. 


A pseudo-Hermitian Hamiltonian $H$ admits a complete eigenbasis $\left\{|\Psi_{\alpha} \rangle\right\}$,~\footnote{Such eigenstates must exist in pairs with complex conjugate eigenvalues~\cite{Mostafazadeh:2001jk}.} while its Hermitian conjugate $H^{\dag}$ admits a corresponding complete eigenbasis $\left\{|\overline{\Psi}_{\alpha} \rangle \right\}$:
\begin{align}
H |\Psi_{\alpha} \rangle = E_{\alpha} |\Psi_{\alpha} \rangle \, , 
\quad
H^{\dagger}|\overline{\Psi}_{\alpha} \rangle = E_{\alpha} |\overline{\Psi}_{\alpha} \rangle \, ,
\label{eq:eigen_state_H_1}
\end{align}
where one single Greek letter, such as $\alpha$, collectively denotes all relevant quantum numbers, including both continuous and discrete degrees of freedom. 
The eigenstates $|\Psi_{\alpha} \rangle$ and $|\overline{\Psi}_{\alpha} \rangle$ are generally distinct, but they share the same real eigenvalue $E_{\alpha} \in \mathbb{R}$, corresponding to $H$ and $H^{\dag}$ respectively.
The requirement that the Hamiltonian $H$ admits a complete set of eigenvectors is fundamentally rooted in the principles of quantum measurement theory. If this condition is not satisfied, then there exist states that have zero overlap with all energy eigenstates. As a result, the total probability of obtaining any energy value in a measurement is identically zero for such states. That is, no energy measurement can yield a meaningful outcome on these states, implying they are physically unpreparable.
The two sets of complete eigenbases $\left\{|\Psi_{\alpha} \rangle, |\overline{\Psi}_{\alpha} \rangle \right\}$ form a biorthonormal system
\begin{align}
\langle \overline{\Psi}_{\alpha'}|\Psi_{\alpha} \rangle = \delta_{\Pi}(\alpha'-\alpha) \, ,
\label{eq:inner_prod_bi_1}
\end{align}
where $\delta_{\Pi}(\alpha'-\alpha)$ stands for a generalized delta function, constructed as the sum of products of Dirac delta functions and Kronecker deltas, corresponding to the continuous and discrete quantum numbers respectively, so that we formally have the integral identity
\begin{align}
\int d\Pi_{\alpha} \,  \delta_{\Pi}(\alpha'-\alpha) = 1 \, .
\end{align}
Therefore, the complete biorthonormalization relation in eq.~\eqref{eq:inner_prod_bi_1} directly yields the completeness relation for the entire Hilbert space:
\begin{align}
\int d\Pi_{\alpha} \, |\overline{\Psi}_{\alpha} \rangle \langle \Psi_{\alpha}| = \int d\Pi_{\alpha} \, |\Psi_{\alpha} \rangle \langle \overline{\Psi}_{\alpha}| = \boldsymbol{1} 
\, .
\label{eq:complete_rel_1}
\end{align}
If a pseudo-Hermitian Hamiltonian $H$ has a complete set of eigenstates associated with an entirely real spectrum, the canonical metric operator $\eta_{0}$ must be positive-definite~\cite{Mostafazadeh:2001nr,Mostafazadeh:2004mx}, i.e., $\eta_{0} \rightarrow \hat{\eta}$, emphasizing its essential role in defining a positive-definite inner product. The pseudo-Hermitian condition in eq.~\eqref{eq:pseudo_H_2} is then promoted to 
\begin{equation}
H^{\#} = \hat{\eta}^{-1}H^{\dag}\hat{\eta} = H \, .
\label{eq:pseudo_H_+}
\end{equation}
In general, for a non-Hermitian operator $H$, the eigenstates $|\Psi_{\alpha} \rangle$ for $H$ and $|\overline{\Psi}_{\alpha} \rangle$ for $H^{\dag}$ are independent.
However, for pseudo-Hermitian Hamiltonians satisfying eq.~\eqref{eq:pseudo_H_+}, these two sets of eigenstates are rather related by a unitary transformation, i.e., the metric operator $\hat{\eta}$. 
To make this relation explicit, we act with $H^{\dag}$ on the state $\hat{\eta}|\Psi_{\alpha} \rangle$
\begin{align}
H^{\dag} \hat{\eta}|\Psi_{\alpha} \rangle
= \hat{\eta} H |\Psi_{\alpha} \rangle
= E_{\alpha} \hat{\eta} |\Psi_{\alpha} \rangle \, ,
\label{eq:H_dag_eta_a_1}
\end{align}
where we have inserted eq.~\eqref{eq:pseudo_H_+}. This shows that $\hat{\eta}|\Psi_{\alpha} \rangle$ is an eigenstate of $H^{\dag}$ with eigenvalue $E_{\alpha}$. After a suitable rescaling of $\hat{\eta}$,~\footnote{If necessary, except for rescaling, $\hat{\eta}$ may be also combined with an additional transformation that maps $|\Psi_{\alpha} \rangle$ to a linear combination of degenerate eigenstates with the same eigenvalue $E_{\alpha}$. Such a transformation clearly commutes with $H$ and thus preserves the pseudo-Hermitian condition.} we therefore identify~\cite{Mostafazadeh:2001jk,Das:2011zza}
\begin{align}
|\overline{\Psi}_{\alpha} \rangle = \hat{\eta}|\Psi_{\alpha} \rangle \, .
\label{eq:H_dag_eta_+}
\end{align}
The self-consistency of this identification is guaranteed by the positivity of the $\hat{\eta}$-inner product and the biorthonormalization relation~\eqref{eq:inner_prod_bi_1}, which implies the $\hat{\eta}$-orthonormalization relation
\begin{align}
\langle \Psi_{\alpha'}|\Psi_{\alpha} \rangle_{\hat{\eta}} = \delta_{\Pi}(\alpha'-\alpha) \, ,
\label{eq:inner_prod_eta_1}
\end{align}
and the $\hat{\eta}$-completeness relation, obtained directly from from eq.~\eqref{eq:complete_rel_1}
\begin{align}
\int d\Pi_{\alpha} \, \hat{\eta} |\Psi_{\alpha} \rangle \langle \Psi_{\alpha}| = \int d\Pi_{\alpha} \, |\Psi_{\alpha} \rangle \langle \Psi_{\alpha}| \hat{\eta} = \boldsymbol{1} 
\, .
\label{eq:complete_rel_eta_1}
\end{align}


Since we work within the relativistic quantum mechanical framework, the energy eigenstates $|\Psi_{\alpha} \rangle$~\eqref{eq:eigen_state_H_1} are generally multi-particle states, furnishing generalized representations of the Poincar\'{e} group introduced in section~\ref{sec:rep_Poin_pH}. The components of the energy-momentum operator $P^{\mu}$, which commute with each other and serve as conserved charges (see section~\ref{sec:rep_Poin_pH}), allow us to construct a basis from common eigenstates of energy-momentum and spin:~\footnote{Additional quantum numbers may be included when necessary. However, for clarity, we focus here on the standard quantum numbers: momentum, spin/helicity, and species label.}
\begin{equation}
|\Psi_{\alpha} \rangle=|\Psi_{\p_{1},\sigma_{1},n_{1};\p_{2},\sigma_{2},n_{2};\cdots}\rangle\,,
\end{equation}
where each particle is labeled by its three-momentum $\p_{i}$, species label $n_{i}$, and spin projection $\sigma_{i}$ (for massive particles) or helicity (for massless particles). 
For such a multi-particle state, its  total energy is simply the sum of the energies of the one-particle energies
\begin{equation}
E_{\alpha} = E_{\p_{1}} + E_{\p_{2}} +  \cdots \, ,
\end{equation}
with no interaction terms, and each individual energy $E_{\p_{i}}$ satisfying the on-shell dispersion relation $E_{\p_{i}} = \sqrt{|\p_{i}|^{2}+m_{i}^{2}}$, where $m_{i}$ is the mass of the particle labeled by $i$.
The true vacuum $|\Omega \rangle$ is assumed to be Lorentz-invariant and to have zero energy (after a zero-point shift) and momentum.
In a typical scattering process, the system is prepared at the early time limit $t \rightarrow -\infty$ with widely separated, non-interacting incoming particles, and evolves to a configuration at $t \rightarrow +\infty$ where the outgoing particles have again ceased to interact and are asymptotically free. It is this asymptotic scenario that justifies the free-like summation of energies shown above. A more detailed analysis of asymptotic behavior will be provided in section~\ref{sec:scatter}.
In this notation, we adopt the relativistic $\hat{\eta}$-orthonormalization relation~\eqref{eq:inner_prod_eta_1} to the following explicit form:
\begin{align}
\langle \Psi_{_{\p'_{1},\sigma'_{1},n'_{1};\p'_{2},\sigma'_{2},n'_{2};\cdots}}|\Psi_{\p_{1},\sigma_{1},n_{1};\p_{2},\sigma_{2},n_{2};\cdots} \rangle_{\hat{\eta}}
&=(2\pi)^{3}2E_{\p_{1}}\delta^{(3)}(\p'_{1}-\p_{1})\delta_{\sigma'_{1} \sigma_{1}} \delta_{n'_{1} n_{1}}
\nonumber \\
&\quad  \times (2\pi)^{3}2E_{\p_{2}}\delta^{(3)}(\p'_{2}-\p_{2})\delta_{\sigma'_{2} \sigma_{2}} \delta_{n'_{2} n_{2}}
\nonumber \\
&\quad  \times  \cdots
\nonumber \\
&\quad  \pm  \text{permutations}
\, ,
\end{align}
where the final term `permutations' accounts for all possible permutations of identical particles among the species sets $\{n_{i}\}$ and $\{n'_{i}\}$ that contribute non-vanishing overlap. The overall sign is determined by the exchange symmetry of the wavefunction: It is $+1$ for bosons (i.e., integer spin) and $-1$ for odd permutations involving fermions (i.e., half-integer spin), in accordance with the spin-statistics theorem.
The right-hand side of this relativistic normalization relation provides an explicit realization of the relativistic delta function $\delta_{\Pi}(\alpha'-\alpha)$ and its corresponding integration measure $d\Pi_{\alpha}$ introduced in eq.~\eqref{eq:inner_prod_bi_1}, thereby clarifying their concrete meaning in the context of multi-particle states labeled by momenta, spin, and species indices.
Therefore, the formal $\hat{\eta}$-completeness relation~\eqref{eq:complete_rel_eta_1} takes the explicit form
\begin{align}
\boldsymbol{1} &= \sum_{N} \sum_{n_{1} \sigma_{1} \cdots n_{N} \sigma_{N}} 
\int d\Pi_{N} \, 
|\Psi_{\p_{1},\sigma_{1},n_{1} \cdots \p_{N},\sigma_{N},n_{N}} \rangle\langle \Psi_{\p_{1},\sigma_{1},n_{1} \cdots \p_{N},\sigma_{N},n_{N}} |\hat{\eta}
\nonumber \\
&= \sum_{N} \sum_{n_{1} \sigma_{1} \cdots n_{N} \sigma_{N}} 
\int d\Pi_{N} \, 
\hat{\eta} |\Psi_{\p_{1},\sigma_{1},n_{1} \cdots \p_{N},\sigma_{N},n_{N}} \rangle\langle \Psi_{\p_{1},\sigma_{1},n_{1} \cdots \p_{N},\sigma_{N},n_{N}} |
\, ,
\label{eq:complete_rel_eta_2}
\end{align}
where
\begin{equation}
d\Pi_{N} \equiv \prod_{i=1}^{N} \frac{d^{3}p_{i}}{(2\pi)^{3}}\frac{1}{2E_{\p_{i}}} \, ,
\end{equation}
is the $N$-body Lorentz-invariant phase space differential.
For the completeness relation to be valid, the summation must run over all physically distinct configurations allowed by the theory. That is, we include only configurations that do not differ only by the exchange of identical particles.

\section{Scattering theory}
\label{sec:scatter}




By definition, a scattering process is a process where the incoming and outgoing particles are prepared and measured long before $t\rightarrow-\infty$ and after the interaction $t\rightarrow+\infty$, respectively. 
In a brief practical view, particles are initially prepared in the distant past, in regions where the interaction potential is negligible. They propagate as free particles until they encounter the interaction region, where their evolution becomes highly nontrivial: particles may reflect, refract, and interfere in complex ways. Eventually, fragments of these particles emerge from the interaction zone and resume free propagation, flying off to spatial infinity. At asymptotically late times, the outgoing particles once again behave as free states. Thus, the central task in scattering theory is to relate the free motion in the distant past to the free motion in the distant future.
This paradigm should be applied to both Hermitian and pseudo-Hermitian scattering systems.

\subsection{Asymptotic states}
\label{sec:in_out}

The key working assumption in scattering calculations is that all interactions are confined in a finite time interval $-T<t<T$. This assumption not only reflects the practical setup of real collider experiments, where interactions occur over a localized region in space and time, but also serves a crucial theoretical purpose: it renders the problem well-defined. 
If interactions were present at all times, it would be impossible to consistently prepare initial states at $t\rightarrow-\infty$ or to identify well-defined final states at $t\rightarrow+\infty$. 
In the absence of interactions at asymptotic times, the initial and final states can be described as on-shell one-particle states of given momenta, known as \textit{asymptotic states}.
It is important to emphasize how this asymptotic structure is formulated.
To preserve manifest Lorentz invariance, we adopt the Heisenberg picture, in which quantum states are time-independent and encode the entire spacetime history of a system. This is distinct from the Schr\"{o}dinger picture, where states evolve in time while operators generally remain fixed, as discussed in section~\ref{sec:pic_T}. 
Therefore, we do not treat `in' and `out' states as captures of time-dependent states at time limits.
In particular, in scattering theory, the pseudo-Hermitian full Hamiltonian $H$ admits two distinct sets of complete biorthonormal eigenbases: $\{|\Psi_{\alpha}^{+}\rangle, |\overline{\Psi}_{\alpha}^{+} \rangle\}$ and $\{|\Psi_{\alpha}^{-}\rangle, |\overline{\Psi}_{\alpha}^{-} \rangle\}$, corresponding to the in and out states respectively. 
Any in state can be expanded as a superposition of out states, and vice versa.

To make this scattering structure concrete, we suppose to decompose the full Hamiltonian $H$ into two parts:
\begin{equation}
H=H_{0}+V \, . 
\label{eq:scatter_H_1}
\end{equation}
In order to realize the asymptotic free particles, $H_{0}$ is the free Hamiltonian, governing the free motion at time limits where all interactions — including pseudo-Hermitian effects — are turned off.
Thus, we require $H_{0}$ to be Hermitian as in conventional Hermitian quantum mechanics, and to admit an eigenbasis $\{|\Phi_{\alpha} \rangle\}$ with a corresponding real spectrum $E_{\alpha}$
\begin{align}
H_{0}|\Phi_{\alpha} \rangle&=E_{\alpha} |\Phi_{\alpha} \rangle \, ,
\quad 
E_{\alpha} \in \mathbb{R} \, ,
\label{eq:spec_H0_1}
\end{align}
which form an orthonormal and complete basis~\footnote{Note that, since $H_{0}$ is Hermitian, its eigenbasis $\{|\Phi_{\alpha} \rangle\}$ is orthonormalized with respect to the Dirac-inner product, rather than the $\eta$-inner product.}
\begin{equation}
\langle \Phi_{\alpha'} |\Phi_{\alpha} \rangle=\delta_{\Pi}(\alpha'-\alpha) \, ,
\quad
\int d\Pi_{\alpha} \, |\Phi_{\alpha} \rangle\langle\Phi_{\alpha} |=\boldsymbol{1} \, .
 \label{eq:ortho_complete_free_1}
\end{equation}
$|\Phi_{\alpha} \rangle$ (i.e., asymptotic states) effectively coincide with the physical in states $|\Psi_{\alpha}^{+}\rangle$ in the far past and out states $|\Psi_{\alpha}^{-}\rangle$ in the far future, which are eigenstates of the full Hamiltonian $H$.
The free one-particle state $|\Phi_{\p,\sigma,n} \rangle$ can be generated by the associated creation operator $a^{\dag}(\p,\sigma,n)$ acting once on the vacuum state $|\Phi_{0} \rangle$~\footnote{The vacuum state is normalized dimensionlessly as $\langle \Phi_{0}|\Phi_{0} \rangle = 1$.} 
\begin{equation}
|\Phi_{\p,\sigma,n} \rangle \equiv \sqrt{2 E_{\p,\sigma,n}} \, a^{\dag}(\p,\sigma,n) |\Phi_{0} \rangle \, ,
\label{eq:1p_gen_1}
\end{equation}
where $E_{\p,\sigma,n} = \sqrt{|\p|^{2}+m_{n}^{2}}$ satisfies the on-shell condition. The creation and annihilation operators obey the canonical commutation (for bosons) or anticommutation (for fermions) relations:
\begin{align}
\bigl[a(\p,\sigma,n),a^{\dag}(\p^{\prime},\sigma',n')\bigr]_{\mp} 
&= (2\pi)^{3} \delta^{(3)} \left( \p^{\prime} - \p \right) \delta_{\sigma^{\prime} \sigma} \delta_{n' n} \, ,
\label{eq:can_comm_1} \\
\bigl[a(\p,\sigma,n), a(\p^{\prime},\sigma',n')\bigr]_{\mp}
&= \bigl[a^{\dag}(\p,\sigma,n), a^{\dag}(\p^{\prime},\sigma',n')\bigr]_{\mp}
=0 \, ,
\label{eq:can_comm_2}
\end{align}
with the signs $-$ and $+$ indicating a commutator (for bosons) and an anticommutator (for fermions) respectively.
One can clearly verify that these canonical quantization relations~\eqref{eq:can_comm_1}-\eqref{eq:can_comm_2} are compatible with the relativistic orthonormalization of one-particle states given in eq.~\eqref{eq:ortho_complete_free_1}.
$|\Phi_{0} \rangle$ denotes the free vacuum state, which is assumed to be unique, defined such that it is annihilated to zero by any annihilation operator
\begin{equation}
a(\p,\sigma,n) |\Phi_{0} \rangle = 0 \, .
\label{eq:vac_1}
\end{equation}
In particular, the $N$-particle free state can be obtained by successively acting $N$ creation operators on the vacuum: 
\begin{align}
|\Phi_{\alpha} \rangle =
|\Phi_{\p_{1},\sigma_{1},n_{1};\p_{2},\sigma_{2},n_{2};\cdots \p_{N},\sigma_{N},n_{N}}\rangle
= \mathfrak{a}^{\dag}(\alpha) |\Phi_{0} \rangle \, ,
\label{eq:Np_gen_1}
\end{align}
where 
\begin{align}
\mathfrak{a}^{\dag}(\alpha)
\equiv
\left( \prod_{i=1}^{N} \sqrt{2E_{\p_{i}}} \right) a^{\dag}(\p_{1},\sigma_{1},n_{1}) a^{\dag}(\p_{2},\sigma_{2},n_{2}) \cdots a^{\dag}(\p_{N},\sigma_{N},n_{N}) \, ,
\label{eq:Np_gen_2}
\end{align}
is the relativistically normalized creation operator that generates the same physical state as $\prod a^{\dag}(\p_{i},\sigma_{i},n_{i})$, with the appropriate energy factors inserted to ensure Lorentz invariant normalization~\eqref{eq:ortho_complete_free_1}:
\begin{align}
\bigl[\mathfrak{a}(\alpha),\mathfrak{a}^{\dag}(\alpha')\bigr]_{\mp} 
&= \delta_{\Pi}(\alpha'-\alpha) \, ,
\label{eq:can_comm_Lorentz_1} \\
\bigl[\mathfrak{a}(\alpha), \mathfrak{a}(\alpha')\bigr]_{\mp}
&= \bigl[\mathfrak{a}^{\dag}(\alpha), \mathfrak{a}^{\dag}(\alpha')\bigr]_{\mp}
=0 \, .
\label{eq:can_comm_Lorentz_2}
\end{align}

The interaction term $V$ encapsulates nontrivial dynamics that distinguish the full theory from the effective free theory and can often be treated perturbatively in practical calculations. 
It implies that for each free state, i.e., asymptotic state, there is a physical state (i.e., in state) that asymptotically approaches it in the far past, and another one (i.e., out state) that evolves into it in the far future:~\footnote{More precisely, the asymptotic relation in eq.~\eqref{eq:asymp_state_0} should be understood in the distributional sense: it defines a physically meaningful scattering particle state only when applied to smooth superpositions of energy eigenstates, i.e., a genuinely normalizable wavepacket.}
\begin{equation}
\lim_{t\rightarrow\mp\infty} \Big \lVert e^{-iHt} |\Psi_{\alpha}^{\pm}\rangle
- e^{-iH_{0}t} |\Phi_{\alpha} \rangle \Big \rVert = 0 \, ,
\label{eq:asymp_state_0}
\end{equation}
where the norm is defined via the Dirac-inner product, since it is measured in the asymptotic times where the interaction turns off. This construction preserves the consistency of the dimensionality by ensuring that the space spanned by the asymptotic free states matches the dimensionality of the full Hilbert space spanned by the in or out states.
It is important to emphasize that this asymptotic convergence imposes a significant constraint that the pseudo-Hermitian Hamiltonian $H$ must possess an entirely real spectrum, as previously invoked in section~\ref{sec:bi_basis_H}.
To demonstrate this explicitly, suppose that $|\Psi_{\alpha}^{\pm}\rangle$ is an eigenstate of $H$ with a complex eigenvalue $E_{\alpha}$, then the asymptotic norm in eq.~\eqref{eq:asymp_state_0} becomes
\begin{equation}
\lim_{t\rightarrow\mp\infty} \Big \lVert e^{-iHt} |\Psi_{\alpha}^{\pm}\rangle
- e^{-iH_{0}t} |\Phi_{\alpha} \rangle \Big \rVert 
=
\biggl\{
\begin{array}{r c l}
&e^{2|\infty \mathrm{Im} \, E_{\alpha}|} \langle \Psi_{\alpha}^{\pm} | \Psi_{\alpha}^{\pm} \rangle \\
&\langle \Phi_{\alpha} |\Phi_{\alpha} \rangle
\end{array}
\neq 0 \, ,
\label{eq:asymp_state_01}
\end{equation}
which clearly cannot vanish unless the exponential factor induced by the imaginary part of $E_{\alpha}$ is eliminated. That is, the suppression or divergence factor $e^{t \, \mathrm{Im} \, E_{\alpha}}$ due to time evolution $e^{-iHt}$ must be removed by requiring $\mathrm{Im} \, E_{\alpha} = 0$. Hence, a real spectrum of the pseudo-Hermitian full Hamiltonian $H$ is required for a consistent and well-defined scattering theory in pseudo-Hermitian quantum mechanics within this Hilbert space structure.
The in and out states can now be defined as eigenstates of $H$, corresponding to a common real spectrum
\begin{align}
H |\Psi_{\alpha}^{\pm} \rangle = E_{\alpha} |\Psi_{\alpha}^{\pm} \rangle 
\, ,\quad 
E_{\alpha} \in \mathbb{R} \, ,
\label{eq:spec_H_1}
\end{align}
which are $\hat{\eta}$-orthonormalized as eq.~\eqref{eq:inner_prod_eta_1}
\begin{align}
\langle \Psi_{\alpha'}^{\pm} | \Psi_{\alpha}^{\pm} \rangle_{\hat{\eta}} = \delta_{\Pi}(\alpha'-\alpha) \, .
\label{eq:inner_prod_eta_+-}
\end{align}
Since the scattering process is assumed to be adiabatic, the full pseudo-Hermitian Hamiltonian $H$ is taken to share the same energy spectrum as the free Hamiltonian $H_{0}$~\eqref{eq:spec_H_1}. This requires that the masses appearing in $H_{0}$ must coincide with the physical (i.e., experimentally measured) masses, rather than with any bare mass parameters that may appear in $H$, which is also a fundamental principle in conventional Hermitian scattering theory.
Any discrepancy between these two masses must be incorporated into the interaction term $V$, not $H_{0}$. In addition, any relevant bound states present in the spectrum of $H$ must be included in $H_{0}$ as if they were elementary particles. This treatment ensures that all asymptotic degrees of freedom are appropriately accounted for within the free theory, maintaining a consistent and complete description of scattering in the asymptotic regions.
Furthermore, the asymptotic indistinguishability condition in eq.~\eqref{eq:asymp_state_0} maps the in and out states to the free states in an explicit formula as
\begin{equation}
|\Psi_{\alpha}^{\pm}\rangle=\Omega_{\pm}|\Phi_{\alpha} \rangle \, ,
\label{eq:asymp_state_1}
\end{equation}
where
\begin{equation}
\Omega(t) \equiv e^{iHt}e^{-iH_{0}t} \, ,\quad
\Omega_{\pm}=\lim_{t\rightarrow\mp\infty}\Omega(t) \, .
\label{eq:asymp_Omega_1}
\end{equation}
Thus, the operators $\Omega_{+}$ and $\Omega_{-}$ relate the free-particle states to the exact physical solutions of the full pseudo-Hermitian theory in the distant past and the distant future, respectively. 
To clarify the asymptotic relation~\eqref{eq:asymp_state_1} within the adiabatic framework, we would like to express $\Omega_{\pm}$ in terms of the interaction term $V$~\eqref{eq:scatter_H_1}, i.e., the Lippmann-Schwinger equation.
To achieve this, we construct the operator $U(t,t')$ from $\Omega(t)$~\eqref{eq:asymp_Omega_1} as 
\begin{equation}
U(t,t') \equiv  \Omega^{-1}(t) \Omega(t') = e^{iH_{0}t} e^{-iH(t-t')}e^{-iH_{0}t'} \, ,
\label{eq:evo_u_1}
\end{equation}
with the boundary condition $U(t,t)=\boldsymbol{1}$ for $t=t'$, and satisfying the composition law
\begin{equation}
U(t,t') U(t',t'') = U(t,t'') \, .
\end{equation}
In particular, this construction yields the following useful coincidence
\begin{equation}
U(0,t) = \Omega(t) \, ,
\quad
U(0,\mp \infty) = \Omega_{\pm} \, .
\label{eq:evo_u_00}
\end{equation}
The time evolution of $U(t,t')$ can be solved by the differential equation:
\begin{equation}
i\partial_{t} U(t,t') = V_{I}(t)U(t,t') \, , 
\label{eq:U_diff_t_1}
\end{equation}
where 
\begin{align}
V_{I}(t) \equiv e^{iH_{0}t} V e^{-iH_{0}t} 
\, ,
\label{eq:V_I_1}
\end{align}
is the interaction term written in the \textit{interaction picture}, with the reference time taken as $t_{0}=0$.
Integrating both sides of eq.~\eqref{eq:U_diff_t_1} from $t=t'=t_{1}$ to $t=t_{2}$ with , we find
\begin{equation}
U(t_{2},t_{1})
= \boldsymbol{1} - i\int_{t_{1}}^{t_{2}} dt \,  V_{I}(t) U(t,t_{1})
\, .
\label{eq:U_int_t_1}
\end{equation}
Now, setting $t_{2}=0$, taking the limit $t_{1}\rightarrow\mp\infty$, and using the identification in eq.~\eqref{eq:evo_u_00}, we obtain the expression for $\Omega_{\pm}$:
\begin{equation}
\Omega_{\pm} = \boldsymbol{1} - i\int_{\mp \infty}^{0} dt \,  e^{iH_0t} V e^{-iHt} \Omega_{\pm} \, .
\label{eq:asymp_Omega_2}
\end{equation}
To derive the Lippmann-Schwinger equation, it is necessary to rewrite the right-hand side of eq.~\eqref{eq:asymp_Omega_2} purely in terms of the operators $H_{0}$ and $\Omega_{\pm}$. To achieve this, we consider the commutator
\begin{equation}
[H_{0},\Omega(t)]=\Omega(t)V_{I}(t)-V\Omega(t) \, , 
\label{eq:com_H_0_Omega_1} 
\end{equation}
which directly follows from expressing the effective free Hamiltonian $H_{0}$ as the difference between the full Hamiltonian $H$ and the interaction term $V$, as defined in eq.~\eqref{eq:scatter_H_1}. 
In the asymptotic time limits $t\rightarrow\pm\infty$, $V_{I}(\pm\infty)$ must vanish due to the asymptotic framework.~\footnote{In general, an adiabatic function is typically attached to the interaction term to regulate its temporal support, ensuring that the interaction vanishes in the asymptotic past and future~\cite{Coleman:2018mew}. This renders the full theory asymptotically equivalent to a free theory and makes the scattering problem well-defined.} Thus, eq.~\eqref{eq:com_H_0_Omega_1} simplifies at asymptotic regions as
\begin{equation}
H\Omega_{\pm} = \Omega_{\pm}H_{0} \, ,
\label{eq:com_H_0_Omega_2} 
\end{equation}
which directly leads to
\begin{equation}
e^{-iHt}\Omega_{\pm} = \Omega_{\pm}e^{-iH_{0}t} \, . 
\label{eq:com_H_0_Omega_3}
\end{equation}
Moreover, eq.~\eqref{eq:com_H_0_Omega_2} also confirms that $H$ and $H_{0}$ share the same energy spectrum, as we have declared earlier. This identification can now be explicitly verified by
\begin{equation}
E_{\alpha} |\Psi_{\alpha}^{\pm}\rangle
= H|\Psi_{\alpha}^{\pm}\rangle
= H\Omega_{\pm}|\Phi_{\alpha} \rangle 
= \Omega_{\pm} H_{0} |\Phi_{\alpha} \rangle 
= E_{\alpha} \Omega_{\pm} |\Phi_{\alpha} \rangle 
\, .
\label{eq:asymp_state_2}
\end{equation}
Substituting eq.~\eqref{eq:com_H_0_Omega_3} into eq.~\eqref{eq:asymp_Omega_2}, we obtain $\Omega_{\pm}$ in the desired form
\begin{equation}
\Omega_{\pm}= \boldsymbol{1}-i\int_{\mp \infty}^{0}dt \, e^{iH_0t} V\Omega_{\pm}e^{-iH_{0}t} \, ,
\label{eq:asymp_Omega_3}
\end{equation}
which determines $\Omega_{\pm}$ recursively through the interaction term $V$. To first order in perturbation theory, the solution simplifies to
\begin{align}
\Omega_{\pm} = \boldsymbol{1}-i\int_{\mp \infty}^{0}dt \, V_{I}(t) \, ,
\quad
\Omega_{\pm}^{-1} = \boldsymbol{1}+i\int_{\mp \infty}^{0}dt \, V_{I}(t) \, .
\label{eq:asymp_Omega_pert_1}
\end{align}
The in and out states are constructed by acting $\Omega_{\pm}$ on the free state $|\Phi_{\alpha} \rangle$ as formulated in the asymptotic relation~\eqref{eq:asymp_state_1}, so that we obtain
\begin{equation}
|\Psi_{\alpha}^{\pm}\rangle
= |\Phi_{\alpha}\rangle -i\int^{0}_{\mp\infty} dt \,  e^{i(H_{0}-E_{\alpha})t} V \Omega_{\pm} |\Phi_{\alpha}\rangle
\, , 
\label{eq:asymp_Omega_4}
\end{equation}
where $E_\alpha$ is the eigenvalues of $H_0$ corresponding to the eigenstate $|\Phi_{\alpha}\rangle$. 
To regularize the divergent integral in eq.~\eqref{eq:asymp_Omega_4}, we introduce an exponential regulator $e^{\pm \epsilon t}$ with an infinitesimal positive factor $\epsilon > 0$. The regulated expression yields the Lippmann-Schwinger equation
\begin{equation}
|\Psi_{\alpha}^{\pm}\rangle
= |\Phi_{\alpha}\rangle 
+ \frac{V \Omega_{\pm}}{E_{\alpha} - H_{0} \pm i\epsilon} |\Phi_{\alpha}\rangle
\, ,
\label{eq:LS_1}
\end{equation}
or expanding in the free orthonormal basis $\{|\Phi_{\alpha} \rangle\}$~\eqref{eq:ortho_complete_free_1}
\begin{align}
|\Psi_{\alpha}^{\pm}\rangle
= |\Phi_{\alpha}\rangle 
+ \int d\Pi_{\beta} \,  \frac{T_{\beta \alpha}^{\pm}}{E_{\alpha} - E_{\beta} \pm i\epsilon} |\Phi_{\beta} \rangle
\, ,
\quad \text{with} \quad
T_{\beta \alpha}^{\pm} \equiv \langle \Phi_{\beta}| V |\Psi_{\alpha}^{\pm}\rangle
\, .
\label{eq:LS_2}
\end{align}
We can now understand that the labels `$+$' and `$-$' correspond to the `in' state $|\Psi_{\alpha}^{+}\rangle$ and `out' state $|\Psi_{\alpha}^{-}\rangle$ respectively, as they originate from the sign of the regulator $i\epsilon$.






\subsection{Canonical metric operator}
\label{sec:metric_op}

After clarifying the asymptotic framework, we now consider another fundamental building block of the pseudo-Hermitian scattering theory: 
the canonical metric operator $\hat{\eta}$. As noted in section~\ref{sec:bi_basis_H}, $\hat{\eta}$ must be positive-definite, since we require the pseudo-Hermitian scattering Hamiltonian $H$ to admit an entirely real spectrum.  
A systematic method for constructing the canonical metric operator within the general quantum mechanics has been proposed in ref.~\cite{Das:2011zza}. 
However, to describe relativistic scattering with a continuous spectrum, this approach must be extended. 



The in and out states are $\hat{\eta}$-orthonormalized, as shown in eq.~\eqref{eq:inner_prod_eta_+-}, which are just like orthonormalization for the free states but using the standard Dirac-inner product in eq.~\eqref{eq:ortho_complete_free_1}. This essential feature of the scattering theory, which follows from the asymptotic framework, leads to the consistency condition
\begin{equation}
\langle\Phi_{\alpha'}|\Phi_{\alpha}\rangle
=\delta_{\Pi}(\alpha'-\alpha) 
=\langle\Psi^{\pm}_{\alpha'}|\Psi^{\pm}_{\alpha}\rangle_{\hat{\eta}} =\langle\Phi_{\alpha'}|\Omega^{\dag}_{\pm}\hat{\eta}\Omega_{\pm}|\Phi_{\alpha}\rangle
\, ,
\label{eq:eta_+_con_1}
\end{equation}
which holds for all the eigenstates if and only if  
\begin{equation}
\Omega^{\dag}_{\pm}\hat{\eta}\Omega_{\pm}=\mathbf{1} \, ,
\end{equation}
from which we immediately obtain two natural solutions for the canonical metric operator $\hat{\eta}$:
\begin{equation}
\hat{\eta}_{\pm} = \hat{\eta}_{\pm}^{\dag} = \left( \Omega_{\pm}^{-1} \right)^{\dag} \Omega_{\pm}^{-1} 
\, .
\label{eq:eta_+_scatter_1}
\end{equation}
Perturbative solutions of $\hat{\eta}_{\pm}$ to the first order of the interaction term $V$ can be obtained directly form eq.~\eqref{eq:asymp_Omega_pert_1}
\begin{align}
\hat{\eta}_{\pm} = \boldsymbol{1}-i\int_{\mp \infty}^{0}dt \left[ V_{I}^{\dag}(t) - V_{I}(t) \right]
= \boldsymbol{1}-i\int_{\mp \infty}^{0}dt \,  \Delta V_{I}(t) 
\, ,
\label{eq:eta_+_scatter_2}
\end{align}
where $\Delta V \equiv V^{\dag} - V$, and $\Delta V_{I}(t) = e^{iH_{0}t} \Delta V e^{-iH_{0}t}$ denotes the corresponding operator in the interaction picture.
It is important to emphasize that the two metric operators $\hat{\eta}_{+}$ and $\hat{\eta}_{-}$ orthonormalize the in and out eigenstates, respectively, as given in eq.~\eqref{eq:inner_prod_eta_+-}. That is, $\hat{\eta}_{+}$ applies to the in eigenstates $\{ |\Psi_{\alpha}^{+} \rangle \}$, while $\hat{\eta}_{-}$ applies to the out eigenstates $\{ |\Psi_{\alpha}^{-} \rangle \}$. This structure leads to the corresponding completeness relations:
\begin{align}
\int d\Pi_{\alpha} \, \hat{\eta}_{\pm} |\Psi_{\alpha}^{\pm} \rangle \langle \Psi_{\alpha}^{\pm}| 
= \int d\Pi_{\alpha} \, |\Psi_{\alpha}^{\pm} \rangle \langle \Psi_{\alpha}^{\pm}| \hat{\eta}_{\pm} 
= \boldsymbol{1} 
\, .
\label{eq:complete_rel_eta_+-}
\end{align}
The normalization condition~\eqref{eq:eta_+_con_1} also ensures that the metric operator $\hat{\eta}$ constructed in this manner is positive-definite, as required for a pseudo-Hermitian Hamiltonian with a real spectrum, in accordance with the discussion in section~\ref{sec:bi_basis_H}.
One can then directly verify that the Hamiltonian $H$ is indeed pseudo-Hermitian with respect to this $\hat{\eta}$
\begin{align}
\hat{\eta}^{-1} H^{\dag} \hat{\eta}
= \Omega_{\pm} \left( \Omega_{\pm}^{-1} H \Omega_{\pm} \right)^{\dag} \Omega_{\pm}^{-1}
= \Omega_{\pm} H_{0} \Omega_{\pm}^{-1}
= H
\, ,
\label{eq:pseudo_H_scatter_1}
\end{align}
where we have used the Hermiticity of $H_{0}$, i.e., $H_{0} = H_{0}^{\dag}$, and the asymptotic relation~\eqref{eq:com_H_0_Omega_2}.
It is remarkable that the two canonical metric operators $\hat{\eta}_{\pm}$~\eqref{eq:eta_+_scatter_1} will converge and reduce to the identity operator $\mathbf{1}$ in the Hermitian limit, thereby reproducing the standard Dirac-inner product:
\begin{align}
\hat{\eta}_{\pm} \xrightarrow{\  H \rightarrow H^{\dag} \  } \mathbf{1} \, . 
\label{eq:pH2D_1}
\end{align}
This occurs when $\Omega_{\pm}$ become unitary, as expected in the standard Hermitian QFT.
Furthermore, as discussed in section~\ref{sec:rep_Poin_pH}, the asymptotic free region also admits a set of Hermitian operators, momentum $\P_{0}$,~\footnote{Here, the subscript `$0$' indicates the free theory. One should distinguish the three-momentum operator $\P_{0}$ from $P^{0}=H$, which denote the $0$-component of the four-momentum, i.e., the Hamiltonian.} angular momentum $\J_{0}$, and boost generator $\K_{0}$, which together with $H_{0}$, generate the Lie algebra of the Poincar\'{e} group when acting on free-particle states. 
As in almost all known theories, the interaction term $V$ is just to modify the free Hamiltonian $H_{0}$ into the pseudo-Hermitian full Hamiltonian $H$, while preserving the momentum and angular momentum
\begin{align}
\P = \P_{0} \, , 
\quad
\J = \J_{0} \, .
\label{eq:id_PJ_1}
\end{align}
Then, the Poincar\'{e} algebra, particularly the commutation relations
\begin{align}
[ \J_{0}, H_{0} ] = [ \P_{0}, H_{0} ] = 0 \, , 
\label{eq:com_PJ_1}
\end{align}
combined with eqs.~\eqref{eq:poin_lie_26} and \eqref{eq:poin_lie_26_h}, impose the natural constraints on the interaction term:
\begin{align}
[ \J_{0}, V ] = [ \P_{0}, V ] = [ \J_{0}, V^{\dag} ] = [ \P_{0}, V^{\dag} ] = 0 \, ,
\label{eq:con_V_JP_1}
\end{align}
so that
\begin{alignat}{2}
\P \Omega_{\pm} &= \P_{0} \Omega_{\pm} &= \Omega_{\pm} \P_{0} &= \Omega_{\pm} \P \, , 
\label{eq:com_P_Omega_2} \\
\J \Omega_{\pm} &= \J_{0} \Omega_{\pm} &= \Omega_{\pm} \J_{0} &= \Omega_{\pm} \J \, .
\label{eq:com_J_Omega_2} 
\end{alignat}
As a result, one can clearly see that both $\P$ and $\J$ are indeed pseudo-Hermitian with respect to $\hat{\eta}$ defined in eq.~\eqref{eq:eta_+_scatter_1} 
\begin{align}
\hat{\eta}^{-1}\P^{\dag}\hat{\eta} = \P \, ,
\quad
\hat{\eta}^{-1}\J^{\dag}\hat{\eta} = \J \, .
\label{eq:pseudo_PJ_2}
\end{align}
On the other hand, it's not possible to identify the full boost generator $\K$ with its free counterpart $\K_{0}$, since the full Hamiltonian $H$ differs from the free Hamiltonian $H_{0}$ in the presence of (pseudo-Hermitian) interactions. Thus, we introduce a correction $\W$ to the free boost generator:
\begin{equation}
\K = \K_{0} + \W \, . 
\label{eq:scatter_K_1}
\end{equation}
Similar to the commutator of $H_{0}$ and $\Omega (t)$~\eqref{eq:com_H_0_Omega_1}, we can calculate the commutator of $\K_{0}$ and $\Omega (t)$ and obtain:
\begin{align}
[\K_{0},\Omega(t)] =\Omega(t) \W_{I}(t) - \W \Omega(t) 
\, , \quad
\W_{I}(t) \equiv e^{iH_{0}t} \W e^{-iH_{0}t} \, ,
\label{eq:com_K_0_Omega_1} 
\end{align}
where we have made use of the commutation relations derived from eq.~\eqref{eq:poin_lie_27}:
\begin{align}
[\K,e^{iHt}] = t \P e^{iHt} \, ,
\quad
[\K_{0},e^{-iH_{0}t}] = -t \P_{0} e^{-iH_{0}t} = -t \P e^{-iH_{0}t} \, ,
\end{align}
and the momentum relevant terms cancel.
In the asymptotic time limits $t\rightarrow\pm\infty$, $\W_{I}(\pm\infty)$ must vanish due to the asymptotic framework, consistent with the discussion regarding the asymptotic transition between $H_{0}$ and $H$ in eq.~\eqref{eq:com_H_0_Omega_2}. As a result, eq.~\eqref{eq:com_K_0_Omega_1} simplifies at asymptotic regions as
\begin{equation}
\K \Omega_{\pm} = \Omega_{\pm} \K_{0} \, .
\label{eq:com_K_0_Omega_2} 
\end{equation}
One can then directly verify that the boost generator $\K$ is indeed pseudo-Hermitian with respect to $\hat{\eta}$~\eqref{eq:eta_+_scatter_1}
\begin{align}
\hat{\eta}^{-1} \K^{\dag} \hat{\eta}
= \Omega_{\pm} \left( \Omega_{\pm}^{-1} \K \Omega_{\pm} \right)^{\dag} \Omega_{\pm}^{-1}
= \Omega_{\pm} \K_{0} \Omega_{\pm}^{-1}
= \K
\, ,
\label{eq:pseudo_K_2}
\end{align}
where we have used the Hermiticity of $\K_{0}$, i.e., $\K_{0} = \K_{0}^{\dag}$, along with the asymptotic relation~\eqref{eq:com_K_0_Omega_2}.
Combining eqs.~\eqref{eq:pseudo_H_scatter_1}, \eqref{eq:pseudo_PJ_2}, and \eqref{eq:pseudo_K_2} confirms that a consistent pseudo-Hermitian realization of the Poincar\'{e} algebra~\eqref{eq:pseudo_P_1}-\eqref{eq:pseudo_J_1} 
has been successfully established with respect to $\hat{\eta}$ defined in eq.~\eqref{eq:eta_+_scatter_1}, within the scattering framework.


\subsection{The $S$-matrix}
\label{sec:smatrix}




A scattering experiment generally needs to prepare an initial state with a definite particle content in the distant past, as $t \rightarrow -\infty$, described by the in state $|\Psi_{\alpha}^{+}\rangle$ and the corresponding asymptotic (free) state $|\Phi_{\alpha} \rangle$.
One then measures the state in the far future, as $t \rightarrow +\infty$, described by the out state $|\Psi_{\beta}^{-}\rangle$ and the corresponding asymptotic (free) state $|\Phi_{\beta} \rangle$. 
This asymptotic relation is given explicitly in eq.~\eqref{eq:asymp_state_1}.
The probability amplitude for this transition $\alpha \rightarrow \beta$ is formulated by the inner product, known as the $S$-matrix.
In systems governed by a Hermitian Hamiltonian, the elements of the $S$-matrix are given by the Dirac-inner product between the in and out states. 
In contrast, as discussed in section~\ref{sec:pH_QM}, the Dirac-inner product is no longer conserved in a pseudo-Hermitian system.
Instead, the $S$-matrix can be naturally defined via the $\hat{\eta}$-inner product in a pseudo-Hermitian scattering system.
Although one is free to adopt either $\hat{\eta}_{+}$ or $\hat{\eta}_{-}$~\eqref{eq:eta_+_scatter_1}, this choice does not affect the generality of the formalism. However, for definiteness, it is necessary to declare the canonical metric operator initially. 
Physical predictions for observed scattering amplitudes can only be extracted once the metric operator $\hat{\eta}$ has been fixed. Conversely, the observed scattering amplitude constrains the coupling constants in the Lagrangian only after $\hat{\eta}$ has been fully specified, in addition to the standard requirements such as the regularization and the subtraction scheme. 
The choice of $\hat{\eta}$ thus constitutes a necessary and complementary prescription, regarded as equally fundamental as these conventional procedures. 
Meanwhile, the in states $\{ |\Psi_{\alpha}^{+} \rangle \}$ and out states $\{ |\Psi_{\alpha}^{-} \rangle \}$ are orthonormalized with respect to the two canonical metric operators $\hat{\eta}_{+}$ and $\hat{\eta}_{-}$, respectively, as shown in eq.~\eqref{eq:complete_rel_eta_+-}. Hence, one cannot naively construct the $S$-matrix with the $\hat{\eta}_{\pm}$-inner product in the form $\langle\Psi^{-}_{\beta}|\Psi^{+}_{\alpha}\rangle_{\hat{\eta}_{\pm}}$, by directly generalizing the Hermitian QFT definition.

To resolve this normalization mismatch, we introduce a one-to-one map which matches an $\hat{\eta}_{+}$-normalized state to an $\hat{\eta}_{-}$-normalized state, which is formulated as an in-out projector 
\begin{align}
\hat{\phi}_{*}: \  (\hat{\eta}_{+},\mathscr{H}) &\rightarrow (\hat{\eta}_{-},\mathscr{H}) \, ,
\nonumber \\
\hat{\eta}_{+} &= \hat{\phi}_{*}^{\dag} \hat{\eta}_{-} \hat{\phi}_{*} \, ,
\label{eq:io_ope_1}
\end{align}
with its inverse, the out-in projector, defined as
\begin{align}
\hat{\phi}^{*} \equiv \hat{\phi}_{*}^{-1}: \  (\hat{\eta}_{-},\mathscr{H}) &\rightarrow (\hat{\eta}_{+},\mathscr{H}) \, ,
\nonumber \\
\hat{\eta}_{-} &= \hat{\phi}^{*\dag} \hat{\eta}_{+} \hat{\phi}^{*} \, .
\label{eq:oi_ope_1}
\end{align}
Note that in the Hermitian limit, we require the projectors to reduce to the identity operator: 
\begin{align}
\hat{\phi}_{*} \xrightarrow{\  H^{\dag} \rightarrow H \  } \mathbf{1}  \, ,
\quad
\hat{\phi}^{*} \xrightarrow{\  H^{\dag} \rightarrow H \  } \mathbf{1}  \, ,
\end{align}
thereby recovering the familiar Hermitian QFT framework, where the projector is absent and the naive construction of the $S$-matrix applies.
The corresponding state to the in state $ |\Psi_{\alpha}^{+} \rangle$ is defined as
\begin{align}
|\tilde{\Psi}_{\alpha}^{+} \rangle 
\equiv \hat{\phi}_{*} |\Psi_{\alpha}^{+} \rangle 
\, .
\end{align}
The states $\{ |\tilde{\Psi}_{\alpha}^{+} \rangle \}$ are hence $\hat{\eta}_{-}$-orthonormalized due to the metric projection~\eqref{eq:io_ope_1}:
\begin{align}
\langle \tilde{\Psi}_{\alpha'}^{+} | \tilde{\Psi}_{\alpha}^{+} \rangle_{\hat{\eta}_{-}} 
= \langle \Psi_{\alpha'}^{+} |\hat{\phi}_{*}^{\dag} \hat{\eta}_{-} \hat{\phi}_{*} | \Psi_{\alpha}^{+} \rangle
= \langle \Psi_{\alpha'}^{+} | \hat{\eta}_{+} | \Psi_{\alpha}^{+} \rangle
= \delta_{\Pi}(\alpha'-\alpha) \, ,
\end{align}
and form a complete basis satisfying the completeness relation:
\begin{align}
\int d\Pi_{\alpha} \, \hat{\eta}_{-} |\tilde{\Psi}_{\alpha}^{+} \rangle \langle \tilde{\Psi}_{\alpha}^{+}| 
= \int d\Pi_{\alpha} \, |\tilde{\Psi}_{\alpha}^{+} \rangle \langle \tilde{\Psi}_{\alpha}^{+}| \hat{\eta}_{-} 
= \boldsymbol{1} 
\, .
\label{eq:complete_rel_eta_+p}
\end{align}
In analogy, we can also introduce $\{ |\tilde{\Psi}_{\alpha}^{-} \rangle \}$ from the out states $\{ |\Psi_{\alpha}^{-} \rangle \}$, 
\begin{align}
|\tilde{\Psi}_{\alpha}^{-} \rangle 
\equiv \hat{\phi}^{*} |\Psi_{\alpha}^{-} \rangle 
\, ,
\end{align}
which are $\hat{\eta}_{+}$-orthonormalized via the inverse metric projection~\eqref{eq:oi_ope_1}:
\begin{align}
\langle \tilde{\Psi}_{\alpha'}^{-} | \tilde{\Psi}_{\alpha}^{-} \rangle_{\hat{\eta}_{+}} 
= \langle \Psi_{\alpha'}^{-} |\hat{\phi}^{*\dag} \hat{\eta}_{+} \hat{\phi}^{*} | \Psi_{\alpha}^{-} \rangle
= \langle \Psi_{\alpha'}^{-} | \hat{\eta}_{-} | \Psi_{\alpha}^{-} \rangle
= \delta_{\Pi}(\alpha'-\alpha) \, ,
\end{align}
forming another complete basis with the completeness relation
\begin{align}
\int d\Pi_{\alpha} \, \hat{\eta}_{+} |\tilde{\Psi}_{\alpha}^{-} \rangle \langle \tilde{\Psi}_{\alpha}^{-}| 
= \int d\Pi_{\alpha} \, |\tilde{\Psi}_{\alpha}^{-} \rangle \langle \tilde{\Psi}_{\alpha}^{-}| \hat{\eta}_{+} 
= \boldsymbol{1} 
\, .
\label{eq:complete_rel_eta_-p}
\end{align}
With these two dual constructions, the probability of scattering in a physical experiment can be related to the corresponding transition amplitudes with either metric. Using the $\hat{\eta}_{+}$-inner product, the $S$-matrix is written as
\begin{equation}
S_{\beta\alpha}[\hat{\eta}_{+}] \equiv \langle\tilde{\Psi}^{-}_{\beta}|\Psi^{+}_{\alpha}\rangle_{\hat{\eta}_{+}} 
\, ,
\label{eq:s_matrix_pH_+}
\end{equation}
which connects the two complete sets $\{ |\Psi_{\alpha}^{-} \rangle \}$ and $\{ |\tilde{\Psi}_{\alpha}^{+} \rangle \}$.
Alternatively, with the $\hat{\eta}_{-}$-inner product we obtain
\begin{equation}
S_{\beta\alpha}[\hat{\eta}_{-}] \equiv \langle\Psi^{-}_{\beta}|\tilde{\Psi}^{+}_{\alpha}\rangle_{\hat{\eta}_{-}}
\, ,
\label{eq:s_matrix_pH_-}
\end{equation}
which instead connects $\{ |\tilde{\Psi}^{-}_{\beta} \rangle \}$ and $\{ |\Psi_{\alpha}^{+} \rangle \}$.
Furthermore, one can show explicitly that these two definitions are equivalent:
\begin{align}
S_{\beta\alpha}[\hat{\eta}_{-}] = \langle\Psi^{-}_{\beta}|\hat{\eta}_{-} \hat{\phi}_{*}|\Psi^{+}_{\alpha}\rangle
= \langle\Psi^{-}_{\beta}|\hat{\phi}^{*\dag} \hat{\eta}_{+}|\Psi^{+}_{\alpha}\rangle 
= S_{\beta\alpha}[\hat{\eta}_{+}]
\, ,
\end{align}
where the second step follows from the inverse metric projection~\eqref{eq:oi_ope_1}. Therefore, owing to this exact equivalence, we are justified in writing the $S$-matrix in a unified notation,
\begin{align}
S_{\beta\alpha} = S_{\beta\alpha}[\hat{\eta}_{+}] = S_{\beta\alpha}[\hat{\eta}_{-}] \, ,
\end{align}
thereby omitting the explicit metric label. In practical applications, either representation can be adopted, depending on which form is 
technically more convenient.
It is important to note that not only two canonical metric operators $\hat{\eta}_{\pm}$~\eqref{eq:eta_+_scatter_1} 
but also the projectors $\hat{\phi}_{*}$ and $\hat{\phi}^{*}$, converge to the identity operator $\mathbf{1}$ in the Hermitian limit~\eqref{eq:pH2D_1}. As a result, the $S$-matrix reduces to the standard formula via the Dirac-inner product in the Hermitian QFT.
However, since the $S$-matrix connecting two complete orthonormal sets via the same metric, it must be unitary as usual. This can be explicitly verified by a direct calculation using the explicit form~\eqref{eq:s_matrix_pH_-}
\begin{equation}
\int d\Pi_{\beta} \, S^{\dag}_{\gamma\beta} S_{\beta\alpha}
= \int d\Pi_{\beta} \, \langle\tilde{\Psi}^{+}_{\gamma} | \hat{\eta}_{-} |\Psi^{-}_{\beta}\rangle  \langle\Psi^{-}_{\beta} | \hat{\eta}_{-} |\tilde{\Psi}^{+}_{\alpha}\rangle
=\langle\tilde{\Psi}^{+}_{\gamma}|\hat{\eta}_{-}|\tilde{\Psi}^{+}_{\alpha}\rangle
= \delta_{\Pi}(\gamma-\alpha) \, ,
\label{eq:s_uni_1}
\end{equation}
where the $\hat{\eta}_{-}$-completeness relation for out states~\eqref{eq:complete_rel_eta_+-} is inserted in the second step. 
Similarly, the $\hat{\eta}_{-}$-completeness relation for the projected in states~\eqref{eq:complete_rel_eta_+p} gives
\begin{equation}
\int d\Pi_{\beta} \, S_{\gamma\beta} S^{\dag}_{\beta\alpha}
= \delta_{\Pi}(\gamma-\alpha) \, .
\label{eq:s_uni_2}
\end{equation}

Rather than working directly with the formal definition of the $S$-matrix in terms of in and out states, it is generally more practical to deal with the free states.
By applying the asymptotic relation~\eqref{eq:asymp_state_1} along with the explicit form of $\hat{\eta}_{\pm}$~\eqref{eq:eta_+_scatter_1}, 
the $S$-matrix~\eqref{eq:s_matrix_pH_+}-\eqref{eq:s_matrix_pH_-} can be rewritten as
\begin{equation}
S_{\beta\alpha}=\langle\Phi_{\beta}|S|\Phi_{\alpha}\rangle \, ,
\label{eq:s_matrix_pH_2}
\end{equation}
which in turn yields the formula for the $S$-operator in two equivalent forms
\begin{equation}
S = \left( \Omega_{+}^{-1} \hat{\phi}^{*} \Omega_{-} \right)^{\dag}
= \Omega_{-}^{-1} \hat{\phi}_{*} \Omega_{+} \, ,
\label{eq:s_ope_pH_1}
\end{equation}
thereby making it manifest that $S$ is unitary, i.e., $S^{\dag} S = S S^{\dag} = \mathbf{1}$, following directly from the equivalence of the two forms.~\footnote{The unitarity of the $S$-operator can also be demonstrated by employing the metric projection~\eqref{eq:io_ope_1} and the inverse projection~\eqref{eq:oi_ope_1} via a straightforward calculation by inserting the explicit form of $\hat{\eta}_{\pm}$~\eqref{eq:eta_+_scatter_1}.}
We would like to express $S$ entirely in terms of the potential $V_{I}(t)$, which allows us to construct Feynman diagrams in direct analogy with standard QFT and thus induces perturbative calculations. To achieve this, it is necessary to analyze the time evolution in detail and correspondingly extend both the canonical metric and the metric projections to include explicit time dependence.
The canonical metric operator $\hat{\eta}_{\pm}$~\eqref{eq:eta_+_scatter_1} for in and out states can be generalized to a time-dependent metric $\hat{\eta}(t)$ through the evolution operator $\Omega(t)$~\eqref{eq:asymp_Omega_1}
\begin{align}
\hat{\eta}(t) \equiv \left( \Omega^{-1}(t) \right)^{\dag} \Omega^{-1}(t)  \, ,
\label{eq:eta_t_scatter_1}
\end{align}
while the metric projector $\hat{\phi}_{*}$~\eqref{eq:io_ope_1} is naturally generalized to $\hat{\phi}_{*}(t,t')$, connecting metric operators $\hat{\eta}(t)$ and $\hat{\eta}(t')$ at two time points, defined via
\begin{align}
\hat{\phi}_{*}^{\dag}(t,t') \hat{\eta}(t) \hat{\phi}_{*}(t,t') = \hat{\eta}(t') \, .
\label{eq:tt_ope_1}
\end{align}
This construction immediately implies the existence of the inverse projector,
\begin{align}
\hat{\phi}^{*}(t,t') \equiv \hat{\phi}_{*}(t',t) = \hat{\phi}_{*}^{-1}(t,t') \, ,
\label{eq:tt_ope_2}
\end{align}
and, in particular, guarantees the initial condition at coincident times,
\begin{align}
\hat{\phi}_{*}(t,t) = \hat{\phi}^{*}(t,t) = \mathbf{1} \, ,
\label{eq:tt_ope_3}
\end{align}
together with the composition rule for sequential evolution,
\begin{align}
\hat{\phi}_{*}(t,t'') = \hat{\phi}_{*}(t,t') \hat{\phi}_{*}(t',t'') \, .
\label{eq:tt_ope_comp_1}
\end{align}
In addition, we require $\phi_{*}(t,t') \rightarrow \mathbf{1}$ in the Hermitian limit $H^{\dag} \rightarrow H$, so that the construction~\eqref{eq:s_ope_pH_1} reduces smoothly to the standard Hermitian $S$-operator.
To reformulate the $S$-operator as a time limit, we introduce a modified evolution operator from $U(t,t')$~\eqref{eq:evo_u_1}
\begin{equation}
\tilde{U}(t,t') \equiv  \Omega^{-1}(t) \hat{\phi}_{*}(t,t') \Omega(t') \, .
\label{eq:evo_u_t_1}
\end{equation}
Inserting eq.~\eqref{eq:eta_t_scatter_1} to eq.~\eqref{eq:tt_ope_1}, we find that $\tilde{U}(t,t')$ is unitary, i.e.,
\begin{equation}
\tilde{U}^{\dag}(t,t') \tilde{U}(t,t') = \tilde{U}(t,t') \tilde{U}^{\dag}(t,t') = \mathbf{1} \, ,
\end{equation}
while eq.~\eqref{eq:tt_ope_2} ensures the inverse evolution property
\begin{equation}
\tilde{U}(t',t) = \tilde{U}^{-1}(t,t') = \tilde{U}^{\dag}(t,t') \, .
\end{equation}
Moreover, $\tilde{U}(t,t')$ satisfies the composition law due to the composition property of the projectors~\eqref{eq:tt_ope_1}
\begin{align}
\tilde{U}(t,t'') = \tilde{U}(t,t') \tilde{U}(t',t'') \, ,
\label{eq:tt_ope_comp_2}
\end{align}
and eq.~\eqref{eq:tt_ope_3} implies the initial condition for $t=t'$
\begin{equation}
\tilde{U}(t,t) = \mathbf{1} \, .
\end{equation}
Hence, the $S$-operator~\eqref{eq:s_ope_pH_1}, which was previously defined only at the asymptotic time limits, can now be understood as the time-evolution limit of the modified evolution operator $\tilde{U}(t,t')$:
\begin{align}
S = \tilde{U}(+\infty,-\infty) \, .
\label{eq:s_ope_pH_U}
\end{align}
The next step is to construct a concrete solution for $\hat{\phi}_{*}(t,t')$~\eqref{eq:tt_ope_1}. To this end, notice that the time-dependent projection in eq.~\eqref{eq:tt_ope_1} satisfies the differential equation
\begin{align}
i\partial_{t} \left[\left( \Omega^{-1}(t) \hat{\phi}_{*}(t,t') \right)^{\dag} \Omega^{-1}(t) \hat{\phi}_{*}(t,t') \right]= 0
\, ,
\label{eq:evo_PH_0}
\end{align}
and $\hat{\phi}_{*}(t,t')$ is the unique solution, determined by the initial condition~\eqref{eq:tt_ope_3} and the Hermitian limit. Thus, we reduce eq.~\eqref{eq:evo_PH_0} into the evolution equation involving the Hermitianized interaction potential, 
\begin{align}
i\partial_{t} \left[\Omega^{-1}(t) \hat{\phi}_{*}(t,t') \right]
= \hat{V}_{I}(t) \left[\Omega^{-1}(t) \hat{\phi}_{*}(t,t') \right]
\, ,
\label{eq:evo_PH_1}
\end{align}
where 
\begin{align}
\hat{V}_{I}(t) \equiv \dfrac{1}{2} \left[V_{I}(t) + V_{I}^{\dag}(t) \right] 
\, ,
\label{eq:V_I_H}
\end{align}
is the Hermitianized potential, which naturally reduces to the potential $V_{I}(t)$~\eqref{eq:V_I_1} in the Hermitian limit
\begin{align}
\hat{V}_{I}(t) \xrightarrow{\  H^{\dag} \rightarrow H \  } V_{I}(t)  \, .
\end{align}
The solution of eq.~\eqref{eq:evo_PH_1} for $\Omega^{-1}(t) \hat{\phi}_{*}(t,t')$ would take our desired form of the Dyson series and can be expressed as a time-ordered exponential
\begin{align}
\Omega^{-1}(t) \hat{\phi}_{*}(t,t') = \mathcal{T} \left\{ \exp \left[ -i \int_{t'}^{t} dt'' \, \hat{V}_{I}(t'') \right] \right\} \Omega^{-1}(t') \, ,
\quad (t>t') \, ,
\end{align}
where $\mathcal{T} \{ \}$ refers to a time-ordered product. It immediately leads to
\begin{align}
\hat{\phi}_{*}(t,t') = \Omega(t) \mathcal{T} \left\{ \exp \left[ -i \int_{t'}^{t} dt'' \, \hat{V}_{I}(t'') \right] \right\} \Omega^{-1}(t') \, ,
\quad (t>t') \, ,
\label{eq:tt_ope_ex}
\end{align}
and consequently yields the explicit expression for the modified evolution operator $\tilde{U}(t,t')$ by the Hermitianized potential $\hat{V}_{I}(t)$~\eqref{eq:V_I_H}, which also serves as the core of the recursive expansion
\begin{align}
\tilde{U}(t,t') = \mathcal{T} \left\{ \exp \left[ -i \int_{t'}^{t} dt'' \, \hat{V}_{I}(t'') \right] \right\} \, ,
\quad (t>t') \, .
\label{eq:evo_u_t_ex}
\end{align}
Taking the asymptotic limit $t=+\infty$ and $t'=-\infty$, eq.~\eqref{eq:tt_ope_ex} then gives the explicit form of the metric projector~\eqref{eq:io_ope_1}
\begin{align}
\hat{\phi}_{*} = \Omega_{-} \mathcal{T} \left\{ \exp \left[ -i \int_{-\infty}^{+\infty} dt \, \hat{V}_{I}(t) \right] \right\} \Omega_{+}^{-1} \, ,
\label{eq:io_ope_ex}
\end{align}
while eq.~\eqref{eq:evo_u_t_ex} yields the explicit form of $S$-operator~\eqref{eq:s_ope_pH_1}:
\begin{equation}
S = \mathcal{T} \left\{ \exp \left[ -i \int_{-\infty}^{+\infty} dt  \, \hat{V}_{I}(t) \right] \right\} \, ,
\label{eq:s_ope_pH_2}
\end{equation}
which leads to the perturbative Dyson series:
\begin{equation}
S = \mathbf{1} + \sum^{\infty}_{n=1}\frac{(-i)^{n}}{n!}\int dt_{1}\cdots dt_{n} \, \mathcal{T} \left\{ \hat{V}_{I}(t_{1}) \cdots \hat{V}_{I}(t_{n}) \right\} 
\, ,
\label{eq:s_ope_pH_Dyson_1}
\end{equation}
so that the $S$-matrix admits a perturbative expansion in terms of Feynman diagrams, evaluated by the usual Feynman rules but now incorporating additional Hermitian-conjugate contributions arising from the metric projector.
This anomalous feature does not occur in the conventional Hermitian theory, where no distinction exists between the metric operators for the in and out states, both of which converge to the identity operator. In the Hermitian limit, where the metric operators $\hat{\eta}_{\pm} \rightarrow \mathbf{1}$ and the metric projector $\hat{\phi}_{*} \rightarrow \mathbf{1}$ as $H^{\dag} \rightarrow H$, the present pseudo-Hermitian formalism consistently reproduces the results of ordinary Hermitian QFT.






\section{Symmetries of pseudo-Hermitian interactions}
\label{sec:scatter_symmetry}



In this section, we discuss the meaning of symmetry in a pseudo-Hermitian theory in the context of the $S$-matrix. We further identify the conditions that the Hamiltonian must satisfy to preserve such symmetry and compare them with those in conventional Hermitian QFT.

When we claim that a pseudo-Hermitian (scattering) theory admits a certain symmetry, we mean that there exists an associated pseudo-unitary and linear operator $U_{-}(T)$ with respect to the metric $\hat{\eta}_{-}$, i.e.,
\begin{equation}
U_{-}^{\dag}(T) \hat{\eta}_{-} U_{-}(T) = \hat{\eta}_{-} \, ,
\label{eq:sym_uni_gen_1-}
\end{equation}
which leaves the $S$-matrix~\eqref{eq:s_matrix_pH_-} invariant
\begin{equation}
S_{\beta\alpha} 
\rightarrow
\langle U_{-}(T) \Psi^{-}_{\beta}|U_{-}(T) \tilde{\Psi}^{+}_{\alpha}\rangle_{\hat{\eta}_{-}}
= \langle\Psi^{-}_{\beta}|\tilde{\Psi}^{+}_{\alpha}\rangle_{\hat{\eta}_{-}}
= S_{\beta\alpha} 
\, ,
\label{eq:sym_con_uni_1}
\end{equation}
so that all the measurable predictions of the theory remain unchanged.
Although we have employed the $S$-matrix using the metric operator $\hat{\eta}_{-}$ here, one can also equivalently invoke the formulation based on $\hat{\eta}_{+}$ with the associated pseudo-unitary and linear operator $U_{+}(T)$
\begin{equation}
U_{+}^{\dag}(T) \hat{\eta}_{+} U_{+}(T) = \hat{\eta}_{+} \, .
\label{eq:sym_uni_gen_1+}
\end{equation}
Moreover, one can relate $U_{+}(T)$ and $U_{-}(T)$ through the metric projector $\hat{\phi}_{*}$~\eqref{eq:io_ope_1} as
\begin{align}
U_{-}(T) &= \hat{\phi}_{*} U_{+}(T) \hat{\phi}^{*} \, , 
\label{eq:sym_uni_gen_1-1} \\
U_{+}(T) &= \hat{\phi}^{*} U_{-}(T) \hat{\phi}_{*} \, .
\label{eq:sym_uni_gen_1+1}
\end{align}
Alternatively, a symmetry transformation may be realized by a pseudo-antiunitary and antilinear operator $U_{-}(T)$ that preserves the metric operator $\hat{\eta}_{-}$ in the same manner as the pseudo-unitary case in eq.~\eqref{eq:sym_uni_gen_1-}. Such an operator maps the $S$-matrix to its complex conjugate
\begin{equation}
S_{\beta\alpha} 
\rightarrow
\langle U_{-}(T) \Psi^{-}_{\beta}|U_{-}(T) \tilde{\Psi}^{+}_{\alpha}\rangle_{\hat{\eta}_{-}}
= \langle\Psi^{-}_{\beta}|\tilde{\Psi}^{+}_{\alpha}\rangle_{\hat{\eta}_{-}}^{*}
= S_{\beta\alpha}^{*}
\label{eq:sym_con_auni_1}
\, ,
\end{equation}
thereby also preserving all observable quantities, since physical cross sections depend only on $|S_{\beta\alpha}|^2$.
The pseudo-antiunitary case likewise admits the $\hat{\eta}_{+}$ formulation with a corresponding pseudo-antiunitary and antilinear operator $U_{+}(T)$ satisfying eq.~\eqref{eq:sym_uni_gen_1+}, while $U_{+}(T)$ and $U_{-}(T)$ are still related as in eq.~\eqref{eq:sym_uni_gen_1+1}.

For practical purposes, it is convenient to reformulate these two types of symmetry conditions on the $S$-matrix in terms of the $S$-operator and the free asymptotic states as in eq.~\eqref{eq:s_matrix_pH_2}. In this formalism, a symmetry in a pseudo-Hermitian theory should correspond to a unitary and linear operator $U_{0}(T)$ acting on free states that preserve the $S$-matrix as in eq.~\eqref{eq:sym_con_uni_1}
\begin{equation}
S_{\beta\alpha} 
\rightarrow
\langle U_{0}(T) \Phi_{\beta} | S |U_{0}(T) \Phi_{\alpha}\rangle
= \langle \Phi_{\beta} | U_{0}^{\dag}(T) S U_{0}(T) |\Phi_{\alpha}\rangle
= \langle \Phi_{\beta} | S |\Phi_{\alpha}\rangle
= S_{\beta\alpha} 
\, ,
\label{eq:sym_con_uni_21}
\end{equation}
where the symmetry operator $U_{0}(T)$ is required to hold the $S$-operator invariant
\begin{equation}
U_{0}^{\dag}(T) S U_{0}(T) = S \, .
\label{eq:sym_con_uni_22}
\end{equation}
Alternatively, a distinct class of symmetries can be realized through an antiunitary and antilinear operator $U_{0}(T)$ that maps the $S$-matrix to its complex conjugate, as in eq.~\eqref{eq:sym_con_auni_1}:
\begin{align}
S_{\beta\alpha} 
\rightarrow
\langle U_{0}(T) \Phi_{\beta} | S^{\dag} |U_{0}(T) \Phi_{\alpha}\rangle
= \langle \Phi_{\beta} | U_{0}^{\dag}(T) S^{\dag} U_{0}(T) |\Phi_{\alpha}\rangle^{*}
= \langle \Phi_{\beta} | S |\Phi_{\alpha}\rangle^{*}
= S_{\beta\alpha}^{*}
\, ,
\label{eq:sym_con_auni_21}
\end{align}
with the corresponding symmetry condition on the $S$-operator given by
\begin{equation}
U_{0}^{\dag}(T) S U_{0}(T) = S^{\dag} \, .
\label{eq:sym_con_auni_22}
\end{equation}
Once a specific symmetry is selected, $U_{0}(T)$ can be determined conveniently from its action on the free states. The full transformation operators $U_{\pm}(T)$ are in turn obtained from the in and out asymptotic process~\eqref{eq:asymp_state_1}. 
The transformed states $|U_{\pm}(T) \Psi^{\pm}_{\alpha}\rangle$ (or $|U_{\pm}(T) \tilde{\Psi}^{\mp}_{\alpha}\rangle$) and $|U_{0}(T) \Phi_{\alpha}\rangle$ can be expressed as superpositions over the eigenbasis, with coefficients fixed by the imposed symmetry. This construction naturally ensures the covariance of the $S$-matrix through a procedure analogous to that employed in standard Hermitian QFT.

\subsection{Lorentz invariance}


Lorentz symmetry constitutes the fundamental requirement of any relativistic theory.
As discussed in section~\ref{sec:metric_op}, the generators of the Poincar\'{e} group for the full theory are pseudo-Hermitian with respect to the canonical metric operators $\hat{\eta}_{\pm}$~\eqref{eq:eta_+_scatter_1} constructed by the asymptotic operators $\Omega_{\pm}$, which leads to the corresponding pseudo-unitary transformation operators $U(\Lambda,a)$ on the Hilbert space
\begin{equation}
U^{\dag}(\Lambda,a) \hat{\eta}_{\pm} U(\Lambda,a) = \hat{\eta}_{\pm} \, ,
\label{eq:eta_sym}
\end{equation}
where $a^{\mu}$ is a constant four-vector for spacetime translations and $\Lambda^{\mu}_{\ \nu}$ is a constant matrix representing homogeneous Lorentz transformations. 
It is notable that $U(\Lambda,a)$ is pseudo-unitary and linear with respect to both $\hat{\eta}_{+}$ and $\hat{\eta}_{-}$ 
\begin{equation}
U(\Lambda,a) = U_{+}(\Lambda,a) = U_{-}(\Lambda,a) \, ,
\end{equation}
which indicates that the Poincar\'{e} symmetry is realized through a single unified set of pseudo-unitary operators. We hence do not need to distinguish them according to the two metric operators $\hat{\eta}_{\pm}$ as in the general case~\eqref{eq:sym_uni_gen_1-}-\eqref{eq:sym_uni_gen_1+}.
In addition, eqs.~\eqref{eq:sym_uni_gen_1-1}-\eqref{eq:sym_uni_gen_1+1} imply that any Poincar\'{e} transformation $U(\Lambda,a)$ must commute with the metric projector $\hat{\phi}_{*}$~\eqref{eq:io_ope_1}
\begin{equation}
[U(\Lambda,a),\hat{\phi}_{*}]=0 \, .
\label{eq:sym_proj_poin_1}
\end{equation}
This condition is equivalent to requiring that all the Poincar\'{e} generators commute with $\hat{\phi}_{*}$.
By imposing the natural constraints on the interaction term in eq.~\eqref{eq:con_V_JP_1} and making use of eqs.~\eqref{eq:id_PJ_1}-\eqref{eq:com_PJ_1}, we obtain additional commutation relations for the momentum and angular momentum operators:
\begin{align}
[ \P, S ] = [ \J, S ] = [ \P, \Omega_{\pm} ] = [ \J, \Omega_{\pm} ] = 0 \, ,
\label{eq:com_JP_2}
\end{align}
which directly implies that $\P$ and $\J$ commute with the metric projector $\hat{\phi}_{*}$~\eqref{eq:io_ope_ex} 
\begin{align}
[ \P, \hat{\phi}_{*} ] = [ \J, \hat{\phi}_{*} ] = 0 \, .
\label{eq:com_JP_3}
\end{align}
The remaining task is to show that the full Hamiltonian $H$ and the boost generator $\K$ also commute with $\hat{\phi}_{*}$.
From the asymptotic commutation relations given in eqs.~\eqref{eq:com_H_0_Omega_2} and \eqref{eq:com_K_0_Omega_2},
\begin{align}
[ H, \hat{\phi}_{*} ] = [ \K, \hat{\phi}_{*} ] = 0 \, ,
\label{eq:com_HK_3}
\end{align}
can be guaranteed if the free Hamiltonian $H_{0}$ and the free boost generator $\K_{0}$ commute with the $S$-operator~\eqref{eq:s_ope_pH_2}
\begin{align}
[ H_{0}, S ] = [ \K_{0}, S ] = 0 \, .
\label{eq:com_HK_4}
\end{align}
Indeed, this commutation relation necessarily holds because the $S$-operator~\eqref{eq:s_ope_pH_2} for a pseudo-Hermitian theory is exactly identical to that of a Hermitian scattering theory governed by the Hermitianized potential $\hat{V}_{I}(t)$~\eqref{eq:V_I_H}.
Therefore, the $S$-matrix~\eqref{eq:s_matrix_pH_-} in a pseudo-Hermitian theory preserves the Poincar\'{e} invariance and can be explicitly written as
\begin{equation}
S_{\beta\alpha} 
= \langle\Psi^{-}_{\beta}|\tilde{\Psi}^{+}_{\alpha}\rangle_{\hat{\eta}_{-}}
= \langle U(\Lambda,a) \Psi^{-}_{\beta}|U(\Lambda,a) \tilde{\Psi}^{+}_{\alpha}\rangle_{\hat{\eta}_{-}}
\, ,
\label{eq:S_sym_Lorentz_1}
\end{equation}
which is consistent with the familiar Hermitian framework.

On the other hand, the Poincar\'{e} invariance conditions in eqs.~\eqref{eq:com_JP_2} and \eqref{eq:com_HK_4} can be derived not only directly from the original definition of the $S$-matrix~\eqref{eq:S_sym_Lorentz_1} with the $\hat{\eta}$-inner product, but also from its free representation involving the $S$-operator~\eqref{eq:sym_con_uni_21}. It clearly implies that the Poincar\'{e} invariance will hold if the free Poincar\'{e} transformation operator $U_{0}(\Lambda,a)$ satisfies eq.~\eqref{eq:sym_con_uni_22}, i.e.,
\begin{equation}
U_{0}^{\dag}(\Lambda,a) S U_{0}(\Lambda,a) = S \, ,
\label{eq:S_sym_Lorentz_2}
\end{equation}
or expressed in terms of infinitesimal Poincar\'{e} transformations that the $S$-operator commutes with the free Poincar\'{e} generators:
\begin{equation}
[H_{0},S]=[\P_{0},S]=[\J_{0},S]=[\K_{0},S]=0 \, .
\end{equation}
In particular, the spacetime translations are represented on the Hilbert space by
\begin{equation}
U(1,a) = e^{-i P^{\mu} a_{\mu}} \, ,
\label{eq:trans_U_1}
\end{equation}
so that the in and out states transform under translations as
\begin{equation}
U(1,a) |\Psi^{\pm}_{\alpha}\rangle 
= \exp \left[-i a_{\mu} \left( p_{1}^{\mu} + p_{2}^{\mu} +  \cdots \right) \right] |\Psi^{\pm}_{\alpha}\rangle \, ,
\label{eq:trans_state_1}
\end{equation}
where the sum runs over all momenta of the particles in the state. This implies that spacetime translations preserve the quantum numbers while contributing only an additional phase factor determined by the total momentum of the state.
Applying eq.~\eqref{eq:trans_state_1} to eq.~\eqref{eq:S_sym_Lorentz_1}, we obtain the translation invariance of the $S$-matrix
\begin{equation}
S_{\beta\alpha} 
= \exp \left[i a_{\mu} \left( p_{1}^{\prime \mu} + p_{2}^{\prime \mu} +  \cdots - p_{1}^{\mu} - p_{2}^{\mu} -  \cdots \right) \right] S_{\beta\alpha}
\, ,
\label{eq:S_sym_trans_1}
\end{equation}
where primes are used to distinguish final from initial particles. Thus, the $S$-matrix vanishes unless the four-momentum is conserved as expected. We can therefore write the part of the $S$-matrix that represents actual interactions among the particles in the form:
\begin{equation}
S_{\beta\alpha} 
= \delta_{\Pi}(\beta -\alpha) + (2\pi)^{4} \delta^{(4)}(p_{\alpha} - p_{\beta}) \cdot i \mathcal{M}_{\beta\alpha}
\, ,
\label{eq:s_matrix_pH_3}
\end{equation}
where $\mathcal{M}_{\beta\alpha}$ denotes the scattering amplitude due to interactions, and the Dirac delta function $\delta^{(4)}(p_{\alpha} - p_{\beta})$ for the energy and momentum conservation is factored out explicitly.

\subsection{Discrete symmetries $P$, $T$, and $C$}



In addition to the continuous Poincar\'{e} transformations, the symmetries of spatial parity $P$, time reversal $T$, and charge conjugation $C$ are commonly considered as potential symmetries of interactions. Nevertheless, there is no fundamental reason to assume that each of these symmetries is exactly conserved. For instance, parity and charge conjugation are symmetries of the strong and electromagnetic interactions, but are violated by the weak interactions due to their chiral nature. In particular, the $CPT$ theorem plays a central role in the theory governed by a Hermitian Hamiltonian, which states that any local, Lorentz-invariant, unitary field theory must be invariant under the product operation $CPT$. Since pseudo-Hermitian QFTs also respect Lorentz symmetry, we still naturally expect $CPT$ invariance to hold exactly.
In this section, we introduce the basic formalism for discrete symmetries in the context of a pseudo-Hermitian theory, and study their conservation condition together with the associated novel physical consequences. 
\\

\noindent \textit{\textbf{Parity}} \\[1ex]
According to the general condition in eq.~\eqref{eq:sym_con_uni_1}, if spatial parity $(t,\x) \rightarrow (t,-\x)$ is indeed a valid symmetry, there must exist a $\hat{\eta}_{\pm}$-pseudo-unitary and $\hat{\phi}_{*}$-commuting linear operator $P$ that transforms both in and out states as a direct product of one-particle states:~\footnote{Here we restrict to massive particles, while the modification for massless particles is straightforward.}
\begin{equation}
P|\Psi^{\pm}_{\alpha}\rangle = \eta_{n_{1}} \eta_{n_{2}} \cdots |\Psi^{\pm}_{\mathscr{P}\alpha}\rangle \, , 
\label{eq:P_state_1}
\end{equation}
where $\eta_{n_{i}}$ is the intrinsic parity phase of the particle species $n_{i}$, and $\mathscr{P}$ reverses the spatial momentum $\p_{i}$ of each particle in $\alpha$, i.e.,
\begin{equation}
\mathscr{P}\alpha = -\p_{1},\sigma_{1},n_{1};-\p_{2},\sigma_{2},n_{2};\cdots \, .
\end{equation}
Using eq.~\eqref{eq:P_state_1}, the parity invariance of the $S$-matrix~\eqref{eq:sym_con_uni_1} then directly implies the covariance relation
\begin{equation}
S_{\beta\alpha} =
\eta_{n'_{1}}^{*} \eta_{n'_{2}}^{*} \cdots \eta_{n_{1}} \eta_{n_{2}} \cdots  S_{\mathscr{P}\beta \mathscr{P}\alpha} 
\, .
\label{eq:P_Smatrix_1}
\end{equation}
Such a parity operator $P$ satisfying eq.~\eqref{eq:P_state_1} will really exist, if a unitary free parity operator $P_{0}$, which is defined to act in the same way on the free states 
\begin{equation}
P_{0} |\Phi_{\alpha}\rangle = \eta_{n_{1}} \eta_{n_{2}} \cdots |\Phi_{\mathscr{P}\alpha}\rangle \, , 
\label{eq:P_state_0_1}
\end{equation}
commutes with the $S$-operator, as we illustrated in eq.~\eqref{eq:sym_con_uni_22}.
The parity transformation on the creation and annihilation operators~\eqref{eq:1p_gen_1} is then derived directly as
\begin{align}
P_{0} a^{\dag}(\p,\sigma,n) P^{-1}_{0} &= \eta_{n} a^{\dag}(-\p,\sigma,n) \, , 
\label{eq:P_a+_0_1} \\
P_{0} a(\p,\sigma,n) P^{-1}_{0} &= \eta^{*}_{n} a(-\p,\sigma,n) \, ,
\label{eq:P_a_0_1}
\end{align}
where the free vacuum $|\Phi_{0} \rangle$ is $P$-invariant
\begin{equation}
P_{0} |\Phi_{0}\rangle = |\Phi_{0}\rangle \, .
\end{equation}
Furthermore, from either the Lippmann-Schwinger equation~\eqref{eq:LS_1} or the asymptotic relation~\eqref{eq:asymp_state_1},
we can show that it is precisely the operator $P_{0}$ that transforms both in and out states as in eq.~\eqref{eq:P_state_1}, i.e.,
\begin{equation}
P = P_{0} \, ,
\end{equation}
if $P_{0}$ commutes with both the free and (Hermitianized) interacting parts of the Hamiltonian
\begin{align}
P^{-1}_{0} H_{0} P_{0} &=H_{0} \, , 
\label{eq:sym_P_con_01} \\   
P^{-1}_{0} V P_{0} &=V \, ,
\label{eq:sym_P_con_02}
\end{align}
which sufficiently preserves the $S$-operator~\eqref{eq:s_ope_pH_2} as required by the symmetry condition~\eqref{eq:sym_con_uni_22} 
\begin{equation}
P_{0}^{\dag} S P_{0} = S \, .
\end{equation}
The parity conditions~\eqref{eq:sym_P_con_01}-\eqref{eq:sym_P_con_02} further imply that the parity operator $P$ is $\hat{\eta}_{\pm}$-pseudo-unitary 
\begin{align}
P^{\dag} \hat{\eta}_{\pm} P &= \hat{\eta}_{\pm} \, ,
\end{align}
and commutes with the metric projector $\hat{\phi}_{*}$
\begin{equation}
[P,\hat{\phi}_{*}]=0 \, ,
\end{equation}
while eq.~\eqref{eq:P_state_1} can be verified directly from either the Lippmann-Schwinger equation~\eqref{eq:LS_1} or the asymptotic relation~\eqref{eq:asymp_state_1}.
Therefore, the parity operator $P$ is simultaneously unitary and pseudo-unitary, which should be interpreted as the fundamental consistency requirement for implementing parity symmetry in a pseudo-Hermitian framework.
\\

\noindent \textit{\textbf{Time reversal}} \\[1ex] 
Let us now turn to a more tricky discrete Lorentz transformation, the time reversal $(t,\x) \rightarrow (-t,\x)$.
If the time reversal constitutes an exact symmetry, there must exist a corresponding $\hat{\eta}_{-}$-pseudo-antiunitary and antilinear operator $T_{-}$ satisfying
\begin{equation}
T_{-}^{\dag} \hat{\eta}_{-} T_{-} = \hat{\eta}_{-} \, ,
\label{eq:sym_T_1-}
\end{equation}
together with another $\hat{\eta}_{+}$-pseudo-antiunitary and antilinear operator $T_{+}$ 
\begin{equation}
T_{+}^{\dag} \hat{\eta}_{+} T_{+} = \hat{\eta}_{+} \, .
\label{eq:sym_T_1+}
\end{equation}
The two operators $T_{\pm}$ are related through the metric projector $\hat{\phi}_{*}$ as
\begin{align}
T_{-} &= \hat{\phi}_{*} T_{+} \hat{\phi}^{*} \, , 
\label{eq:sym_T_1-1} \\
T_{+} &= \hat{\phi}^{*} T_{-} \hat{\phi}_{*} \, ,
\label{eq:sym_T_1+1}
\end{align}
in accordance with the general pseudo-antiunitary relations introduced in eqs.~\eqref{eq:sym_uni_gen_1-1}-\eqref{eq:sym_uni_gen_1+1}.
The two reversal operators $T_{\pm}$ act on in and out states respectively by reversing momentum and spin while interchanging the in and out sectors:
\begin{align}
T_{\pm}|\Psi^{\pm}_{\alpha}\rangle = \zeta_{n_{1}} (-1)^{j_{1}-\sigma_{1}} \zeta_{n_{2}} (-1)^{j_{2}-\sigma_{2}} \cdots |\tilde{\Psi}^{\mp}_{\mathscr{T}\alpha}\rangle \, , 
\label{eq:T_state_1} 
\end{align} 
where $\mathscr{T}$ reverses the spatial momentum $\p_{i}$ and spin $\sigma_{i}$ of each particle in $\alpha$, i.e.,
\begin{equation}
\mathscr{T}\alpha = -\p_{1},-\sigma_{1},n_{1};-\p_{2},-\sigma_{2},n_{2};\cdots \, .
\end{equation} 
Using eqs.~\eqref{eq:sym_T_1-1}-\eqref{eq:sym_T_1+1}, one also obtains the coupled transformation
\begin{align}
T_{\mp}|\tilde{\Psi}^{\pm}_{\alpha}\rangle = \zeta_{n_{1}} (-1)^{j_{1}-\sigma_{1}} \zeta_{n_{2}} (-1)^{j_{2}-\sigma_{2}} \cdots |\Psi^{\mp}_{\mathscr{T}\alpha}\rangle \, . 
\label{eq:T_state_2} 
\end{align} 
Unlike the parity phase $\eta_{n_{i}}$, the time reversal phase $\zeta_{n_{i}}$ is purely conventional and has no physical significance, since it can always be removed by a phase redefinition of the states.

Inserting eq.~\eqref{eq:T_state_1} to eq.~\eqref{eq:sym_con_auni_1}, we obtain the covariance of the $S$-matrix under the time reversal
\begin{equation}
S_{\beta\alpha} =
\zeta_{n'_{1}} (-1)^{j'_{1}-\sigma'_{1}} \zeta_{n'_{2}} (-1)^{j'_{2}-\sigma'_{2}} \cdots \zeta_{n_{1}}^{*} (-1)^{j_{1}-\sigma_{1}} \zeta_{n_{2}}^{*} (-1)^{j_{2}-\sigma_{2}} \cdots S_{\mathscr{T}\alpha \mathscr{T}\beta} 
\, ,
\label{eq:T_Smatrix_1}
\end{equation}
implying that the transition rate for the process $\alpha \rightarrow \beta$ is identical to that for its time-reversed counterpart $\mathscr{T}\beta \rightarrow \mathscr{T}\alpha$.
As we proposed in eq.~\eqref{eq:sym_con_auni_21}, such pseudo-antiunitary and antilinear operators $T_{\pm}$ will represent an exact symmetry if there is an antiunitary and antilinear operator $T_{0}$, defined to act in the same way on the free states 
\begin{equation}
T_{0} |\Phi_{\alpha}\rangle = \zeta_{n_{1}} (-1)^{j_{1}-\sigma_{1}} \zeta_{n_{2}} (-1)^{j_{2}-\sigma_{2}} \cdots |\Phi_{\mathscr{T}\alpha}\rangle \, , 
\label{eq:T_state_0_1}
\end{equation}
and realize the Hermitian conjugation of the $S$-operator
\begin{equation}
T_{0}^{\dag} S T_{0} = S^{\dag} \, ,
\label{eq:sym_con_T_s}
\end{equation}
as required by eq.~\eqref{eq:sym_con_auni_22}.
The time reversal of the creation and annihilation operators~\eqref{eq:1p_gen_1} can then be derived directly as
\begin{align}
T_{0} a^{\dag}(\p,\sigma,n) T^{-1}_{0} &= \zeta_{n} (-1)^{j-\sigma} a^{\dag}(-\p,-\sigma,n) \, , 
\label{eq:T_a+_0_1} \\
T_{0} a(\p,\sigma,n) T^{-1}_{0} &= \zeta^{*}_{n} (-1)^{j-\sigma} a(-\p,-\sigma,n) \, ,
\label{eq:T_a_0_1}
\end{align}
with the free vacuum $|\Phi_{0} \rangle$ being $T$-invariant
\begin{equation}
T_{0} |\Phi_{0}\rangle = |\Phi_{0}\rangle \, .
\end{equation}
The symmetry condition~\eqref{eq:sym_con_T_s} for time reversal is fulfilled when $T_{0}$ commutes with the free Hamiltonian $H_{0}$ and transforms the interacting part $V$ into its Hermitian conjugate $V^{\dag}$
\begin{align}
T^{-1}_{0} H_{0} T_{0} &= H_{0} \, , 
\\   
T^{-1}_{0} V T_{0} &= V^{\dag} \, ,
\end{align}
which implies
\begin{align}
T_{0}^{\dag} \Omega_{\pm} T_{0} &= \left( \Omega_{\mp}^{-1} \right)^{\dag} \, ,
\\
T_{0}^{\dag} \hat{\eta}_{\pm} T_{0} &= \hat{\eta}_{\mp}^{-1} \, ,
\end{align}
and hence the induced transformations on the metric projectors
\begin{align}
T_{0}^{\dag} \hat{\phi}_{*} T_{0} &= \hat{\phi}_{*}^{\dag} \, ,
\\
T_{0}^{\dag} \hat{\phi}^{*} T_{0} &= \hat{\phi}^{*\dag} \, .
\end{align}
In this case, we can take
\begin{align}
T_{+} &= \hat{\eta}_{+}^{-1} T_{0} \hat{\phi}_{*} \, ,
\label{eq:T_T0_1+} \\
T_{-} &= \hat{\eta}_{-}^{-1} T_{0} \hat{\phi}^{*} \, ,
\label{eq:T_T0_1-}
\end{align}
which directly reproduce the relations between the full time reversal operators $T_{\pm}$ given in eqs.~\eqref{eq:sym_T_1-1}-\eqref{eq:sym_T_1+1}
\begin{align}
\hat{\phi}_{*} T_{+} \hat{\phi}^{*} 
= \hat{\phi}_{*} \hat{\eta}_{+}^{-1} T_{0} \hat{\phi}_{*} \hat{\phi}^{*} 
= \hat{\phi}_{*} \hat{\eta}_{+}^{-1} T_{0}
= \hat{\eta}_{-}^{-1} \hat{\phi}^{*\dag} T_{0}
= \hat{\eta}_{-}^{-1} T_{0} \hat{\phi}^{*}
= T_{-}
\, .
\end{align}
Furthermore, by applying the asymptotic relations~\eqref{eq:asymp_state_1}, one verifies that $T_{\pm}$ do act on the in and out states as described in eqs.~\eqref{eq:T_state_1} and \eqref{eq:T_state_2}
\begin{align}
T_{\mp}|\tilde{\Psi}^{\pm}_{\alpha}\rangle 
= \hat{\eta}_{\mp}^{-1} \left( \Omega_{\mp}^{-1} \right)^{\dag} T_{0} |\Phi_{\alpha}\rangle
= \Omega_{\mp} T_{0} |\Phi_{\alpha}\rangle
= \zeta_{n_{1}} (-1)^{j_{1}-\sigma_{1}} \zeta_{n_{2}} (-1)^{j_{2}-\sigma_{2}} \cdots |\Psi^{\mp}_{\mathscr{T}\alpha}\rangle \, . 
\end{align}

In contrast to the parity transformation, which is implemented by a single pseudo-Hermitian operator $P$, the time reversal transformation is described by two distinct pseudo-antiunitary and antilinear operators $T_{\pm}$, associated with the in and out sectors characterized by the canonical metric $\hat{\eta}_{\pm}$ respectively. These two sectors are connected through the metric projectors $\hat{\phi}_{*}$ and $\hat{\phi}^{*}$, which ensure the proper interchange between in and out states. This feature distinguishes the pseudo-Hermitian formulation from the Hermitian framework, where time reversal directly exchanges the in and out states without any mediation. The metric projectors $\hat{\phi}_{*}$ and $\hat{\phi}^{*}$ further serve to map the free time reversal operator $T_{0}$ to the full operators $T_{\pm}$, thereby maintaining consistency between the free and interacting descriptions. In the Hermitian limit, where the metric operators $\hat{\eta}_{\pm} \rightarrow \mathbf{1}$ and the metric projector $\hat{\phi}_{*} \rightarrow \mathbf{1}$, present pseudo-Hermitian formalism smoothly reduces to the standard Hermitian theory, recovering a single antiunitary time reversal operator with $T_{\pm} \rightarrow T_{0}$.
\\

\noindent \textit{\textbf{Charge conjugation}} \\[1ex] 
Charge conjugation is an internal transformation that changes particles into antiparticles $n \rightarrow n^{c}$ while leaving the momenta and spins unchanged.
If charge conjugation is an exact symmetry, there must exist a $\hat{\eta}_{\pm}$-pseudo-unitary and $\hat{\phi}_{*}$-commuting linear operator $C$ formally acting on both in and out states as
\begin{equation}
C|\Psi^{\pm}_{\alpha}\rangle = \xi_{n_{1}} \xi_{n_{2}} \cdots |\Psi^{\pm}_{\alpha^{c}}\rangle \, , 
\label{eq:C_state_1}
\end{equation}
where $\xi_{n_{i}}$ is the charge conjugation phase of the particle species $n_{i}$, and $\alpha^{c}$ denotes a multi-particle state obtained from $\alpha$ by replacing each particle species with its antiparticle,
\begin{equation}
\alpha^{c} = \p_{1},\sigma_{1},n_{1}^{c};\p_{2},\sigma_{2},n_{2}^{c};\cdots \, .
\end{equation}
From the invariance of the $S$-matrix~\eqref{eq:sym_con_uni_1}, one then obtains the covariance relation under charge conjugation
\begin{equation}
S_{\beta\alpha} =
\xi_{n'_{1}}^{*} \xi_{n'_{2}}^{*} \cdots \xi_{n_{1}} \xi_{n_{2}} \cdots  S_{\beta^{c} \alpha^{c}} 
\, .
\label{eq:C_Smatrix_1}
\end{equation}
A pseudo-unitary operator $C$ satisfying eq.~\eqref{eq:C_state_1} will actually exist, if a unitary operator $C_{0}$, defined to act in the same way on the free states 
\begin{equation}
C_{0} |\Phi_{\alpha}\rangle = \xi_{n_{1}} \xi_{n_{2}} \cdots |\Phi_{\alpha^{c}}\rangle \, , 
\label{eq:C_state_0_1}
\end{equation}
commutes with the $S$-operator, as shown in eq.~\eqref{eq:sym_con_uni_22}.
The charge conjugation on the creation and annihilation operators~\eqref{eq:1p_gen_1} follows directly as
\begin{align}
C_{0} a^{\dag}(\p,\sigma,n) C^{-1}_{0} &= \xi_{n} a^{\dag}(\p,\sigma,n^{c}) \, , 
\label{eq:C_a+_0_1} \\
C_{0} a(\p,\sigma,n) C^{-1}_{0} &= \xi^{*}_{n} a(\p,\sigma,n^{c}) \, ,
\label{eq:C_a_0_1}
\end{align}
while the free vacuum $|\Phi_{0} \rangle$ remains $C$-invariant
\begin{equation}
C_{0} |\Phi_{0}\rangle = |\Phi_{0}\rangle \, .
\end{equation}
In analogy with our discussion of the spatial parity, if $C_{0}$ commutes with both the free and interacting parts of the Hamiltonian
\begin{align}
C^{-1}_{0} H_{0} C_{0} &=H_{0} \, , 
\label{eq:sym_C_con_01} \\   
C^{-1}_{0} V C_{0} &=V \, ,
\label{eq:sym_C_con_02}
\end{align}
then one can identify
\begin{equation}
C = C_{0} \, ,
\end{equation}
indicating that the charge conjugation operator $C$ is both unitary and pseudo-unitary.
The charge conjugation conditions~\eqref{eq:sym_C_con_01}-\eqref{eq:sym_C_con_02} ensure the invariance of the $S$-operator~\eqref{eq:s_ope_pH_2} as required by the symmetry condition~\eqref{eq:sym_con_uni_22} 
\begin{equation}
C_{0}^{\dag} S C_{0} = S \, ,
\label{eq:sym_con_C_s}
\end{equation}
and imply that $C$ is $\hat{\eta}_{\pm}$-pseudo-unitary 
\begin{align}
C^{\dag} \hat{\eta}_{\pm} C &= \hat{\eta}_{\pm} \, ,
\end{align}
as well as commuting with the metric projector $\hat{\phi}_{*}$
\begin{equation}
[C,\hat{\phi}_{*}]=0 \, .
\end{equation}
Charge conjugation on the in and out states~\eqref{eq:C_state_1} can thus be simply verified from either the Lippmann-Schwinger equation~\eqref{eq:LS_1} or the asymptotic relation~\eqref{eq:asymp_state_1}.

Finally, let us review the $CPT$ theorem in the pseudo-Hermitian framework. In general, the combined $CPT$ transformation maps each term of the interaction into its Hermitian conjugate
\begin{align}
[CPT]_{0}^{-1} V [CPT]_{0} = V^{\dag} \, ,
\end{align}
with $[CPT]_{0} \equiv C_{0}P_{0}T_{0}$. In a conventional Hermitian theory, the interaction potential satisfies $V = V^{\dag}$, so that $CPT$ trivially commutes with the interaction. Moreover, in both Hermitian and pseudo-Hermitian settings, $CPT$ always commutes with the free Hamiltonian $H_{0}$. A subtle yet important point is that the $S$-operator in the pseudo-Hermitian formalism is generated not directly by the interaction $V$, but by its Hermitianized counterpart $\hat{V}$~\eqref{eq:s_ope_pH_2}. , Consequently, $CPT$ commutes with $\hat{V}$ and thus realizes the same transformation of the $S$-operator as in the Hermitian case
\begin{align}
[CPT]_{0}^{\dag} S [CPT]_{0} = S^{\dag} \, ,
\end{align}
which establishes the validity of the $CPT$ symmetry in a pseudo-Hermitian (scattering) system. Since $CPT$ is pseudo-antiunitary, it relates the $S$-matrix for an arbitrary process to that of the inverse process, where all the incoming particles are transformed into outgoing antiparticles, and vice versa. Although the $CPT$ symmetry may seem counterintuitive at first glance due to the pseudo-unitary evolution, the technical revision arises nontrivially from the metric projector, which mediates between the in and out sectors, a remarkable structural element absent in the conventional Hermitian QFT.

\section{Conclusion} 
\label{sec:conc}

The pseudo-Hermitian QFT provides a natural framework for generalizing conventional Hermitian QFT. 
In this work, we have systematically established the pseudo-Hermitian formulation of QFT and extended it to scattering processes. A central feature of this framework lies in the presence of distinct metric operators for the in and out sectors, which reflect the pseudo-unitary evolution of the system. The two sectors are bridged by the metric projector, a nontrivial structural element that consistently mediates the transition between them and ensures the overall conservation of probability under pseudo-unitary time evolution. 
We construct the metric projector based on fundamental requirements: mapping between $\hat{\eta}_{\pm}$, and reducing to the identity operator in the Hermitian limit. An explicit solution is obtained by solving the extended differential evolution equation. This construction provides a rigorous foundation for a consistent scattering formalism beyond the conventional Hermitian setting and opens promising directions for further theoretical and phenomenological exploration.

We have reviewed the essential technical ingredients underlying this framework. The pseudo-Hermitian approach introduces modified conjugation and generalized completeness relations, which together define an inner product compatible with pseudo-unitarity. 
The metric projector plays a central role in the asymptotic formalism, ensuring that the $S$-operator remains unitary. 
Moreover, we have established a general symmetry formalism and associated conditions within the pseudo-Hermitian context. Unlike the Hermitian case, a single symmetry typically corresponds to two pseudo-unitary operators associated with the two metrics $\hat{\eta}_{\pm}$ respectively. As concrete examples, we have considered fundamental Poincar\'{e} symmetry, as well as discrete symmetries including spatial parity, time reversal, and charge conjugation. 
We confirm that the $CPT$ theorem remains valid in the pseudo-Hermitian context, confirming that the metric projector is appropriately incorporated.

Despite these developments, several open problems remain.
First, it is crucial to relate the $S$-matrix elements to the $n$-point correlation function [Green's function] by constructing the Lehmann-Symanzik-Zimmermann (LSZ) reduction formula in the pseudo-Hermitian framework. This is nontrivial because Green’s functions are not formulated in an asymptotic setting, so the consistent incorporation of the two metrics and the metric projector must be carefully addressed.
Second, it is necessary to further examine the equivalence between pseudo-Hermitian and Hermitian systems. Although the $S$-operator in a pseudo-Hermitian scattering system remains unitary and is generated by the Hermitianized potential, it is important to examine this equivalence at both the operator and observable levels.
Such studies may clarify whether pseudo-Hermiticity is merely a reformulation of an associated Hermitian system or encodes genuinely new physics.
Finally, if observable deviations between pseudo-Hermitian and Hermitian systems exist, it will be essential to construct measurable quantities and propose experimental setups to detect them. These efforts could establish pseudo-Hermitian QFT as a viable extension of the standard framework, potentially offering new insights into unresolved fundamental questions in the SM.

\appendix

\section*{Appendix}

\section{Notations \& conventions}
\label{app:notations_and_conventions}

Throughout the paper, we use the conventions of ref.~\cite{Peskin:1995ev}.
The 4D Minkowski metric is,
\begin{equation}
g_{\mu \nu} = g^{\mu \nu} = \text{diag}(+1, -1, -1, -1) \, ,
\label{metric_1}
\end{equation}
where $\mu, \nu=0,1, 2, 3$.

The $4\times4$ Dirac matrices are taken in the Weyl representation,
\begin{equation}
\gamma^\mu =
\left[\begin{matrix}
0 & \sigma^\mu \\
\bar{\sigma}^\mu & 0
\end{matrix}\right]
\, ,
\quad \text{with} \quad
\biggl\{
\begin{array}{r c l}
\sigma^\mu &=& \left( \mathds{1}_{2 \times 2}, \sigma^i \right) \, , \\
\bar{\sigma}^\mu &=& \left( \mathds{1}_{2 \times 2}, -\sigma^i \right) \, ,
\end{array}
\label{gamma_1}
\end{equation}
where $\mu=0,1,2,3$ and $\sigma^i$ ($i = 1, 2, 3$) are the three Pauli matrices:
\begin{equation}
\sigma^1 =
\left[\begin{matrix}
0 \quad  & \quad 1  \\
1 \quad  & \quad 0  
\end{matrix}\right] \, ,
\quad
\sigma^2 =
\left[\begin{matrix}
0 \quad  & -i \\
i \quad  & 0
\end{matrix}\right] \, ,
\quad
\sigma^3 =
\left[\begin{matrix}
1 \quad  &  0 \\
0 \quad  &  -1
\end{matrix}\right] \, ,
\end{equation}
and
\begin{align}
\left\{\sigma^{i}, \sigma^{j} \right\} = 2\delta^{ij} \, , \quad i,j = 1 \cdots 3 \, ,
\end{align}
where the Kronecker delta function is
\begin{equation}
\delta^{ij} = 
\biggl\{
\begin{array}{r c l}
&1&, \  i=j\\
&0&, \  i \neq j
\end{array} \, .
\end{equation}
The Levi-Civita tensor $\epsilon_{ijk}$, where $i,j,k = 1 \cdots 3$ is totally antisymmetric, with $\epsilon_{123}=+1$.

One also has the chiral operator,
\begin{equation}
\gamma^5 = i \gamma^0 \gamma^1 \gamma^2 \gamma^3 =
\left[\begin{matrix}
- \mathds{1}_{2 \times 2} & 0 \\
0 & \mathds{1}_{2 \times 2}
\end{matrix}\right] \, ,
\label{gamma_2}
\end{equation}
and the chiral projection operators
\begin{equation}
P_{L,R} \equiv \dfrac{\mathds{1} \mp \gamma^{5}}{2} \, .
\end{equation}

\acknowledgments

We would like to thank Ting-Long Feng for collaboration at the early stage of this work. SZ is supported by the Natural Science Foundation of China under Grant No.12347101, No.2024CDJXY022 and No.CSTB2024YCJH-KYXM0070 at Chongqing University.

\bibliography{Bibliography}
\bibliographystyle{JHEP}


\end{document}